\documentclass[prb,aps,reprint,superscriptaddress]{revtex4-1}

\usepackage{amsmath}
\usepackage{amssymb}
\usepackage{physics}
\usepackage{graphicx}
\graphicspath{./}
\usepackage{hyperref}
\hypersetup{
 unicode=false,			
 pdftoolbar=true,		
 pdfmenubar=true,		
 pdffitwindow=false,		
 pdfstartview={FitH},		
 pdfcreator={pdflatex},		
 pdfproducer={vim},		
 pdfnewwindow=true,		
 colorlinks=true,		
 linktoc=page,			
 linkcolor=blue,		
 citecolor=blue,		
 filecolor=blue,		
 urlcolor=blue			
}
\def\rr{{\bf r}}
\def\d{{\bf d}}

\def\q{{\bf q}}

\begin{document}

\title{DMRG on top of plane-wave Kohn-Sham orbitals: case study of defected  boron nitride}
\date{\today}
\author{Gergely Barcza}
\email{barcza.gergely@wigner.mta.hu}
\affiliation{
Wigner Research Centre for Physics, Budapest, PO. Box 49, H-1525,  Hungary}
\affiliation{J. Heyrovsk\'{y} Institute of Physical Chemistry, Czech Academy of Sciences,  Prague, CZ-18223, Czechia}
\author{Viktor Iv\'ady}
\affiliation{
Wigner Research Centre for Physics, Budapest, PO. Box 49, H-1525,  Hungary}
\affiliation{
Department of Physics, Chemistry and Biology, Link\"oping University,  Link\"oping, SE-581 83, Sweden}
\author{Tibor Szilv\'asi}
\affiliation{Department of Chemical and Biological Engineering, University of Wisconsin-Madison,   Madison, WI-53706, USA}
\author{M\'arton V\"or\"os}
\affiliation{
Material Sciences Division, Argonne National Laboratory,  Lemont, IL-60439,  USA}
\author{Libor Veis}
\affiliation{J. Heyrovsk\'{y} Institute of Physical Chemistry, Czech Academy of Sciences, Prague,  CZ-18223, Czechia}
\author{\'Ad\'am Gali}
\affiliation{
Wigner Research Centre for Physics, Budapest, PO. Box 49, H-1525,  Hungary}
\affiliation{
Department of Atomic Physics, Budapest University of Technology and Economics, Budapest, H-1111,  Hungary}
\author{\"Ors Legeza}
\affiliation{
Wigner Research Centre for Physics, Budapest, PO. Box 49, H-1525,  Hungary}

\begin{abstract}
In this paper, we  analyze the numerical aspects of  the  inherently multi-reference density matrix renormalization group (DMRG) calculations on  top of the periodic  Kohn-Sham density functional theory (DFT) using the complete active space (CAS) approach.
Following the technical outline related to the  computation of the Hamiltonian matrix elements  and to the construction of the active space, we illustrate the potential of the framework by studying the vertical many-body  energy spectrum of  hexagonal boron nitride (hBN) nano-flakes  embedding a single boron vacancy point defect  with prominent  multi-reference character.
We investigate the consistency of the DMRG energy spectrum  from the perspective of sample size, basis size, and active space selection protocol.
Results obtained from standard quantum chemical atom-centered basis calculations  and plane-wave  based counterparts show  excellent agreement.
Furthermore, we also discuss the spectrum of the periodic sheet which is in good agreement with extrapolated data of finite clusters.  
These results pave the way toward applying DMRG method in extended correlated solid state systems, such as point qubit in wide band gap semiconductors.
\end{abstract}

\maketitle

\section{Introduction}
By discovering a plethora of color centres in hexagonal boron nitride (hBN) with various point defect complexes in the last years,~\cite{Aharonovich2015, Tran-2016, WrachtruphBN2016, FuchshBN2016, Bassett2017,    ProsciaOptica18, BecherhBN2019} hBN is becoming a focus of interest.
In particular, additionally to favourable optical properties, point defects in hBN with spin   could possibly implement  quantum bits controllable by optically detected magnetic resonance which signal has been recently reported at room temperature.~\cite{DyakonovArXiv2019, WrachtrupArXiv2019}

The standard theoretical approach to study such defects is based on first-principles methods,~\cite{Orellana-2001,Piquini-2005,Si-2007,Bockstedte2011hBN,Huang-2012,Zhang-2013,Cheng-2017,MFord2017,
WestonPhysRevB2018,PlenioACS2018,Sajid2018,Reimers-2018,Sajid-2019b} in particular on  the Kohn-Sham density functional theory (KS-DFT),~\cite{Kohn1965} which established itself as one of the most prevalent numerical methods both in material science and theoretical chemistry offering  balanced compromise between accuracy and computational costs.  
Despite its immense success attained in the field of weakly correlated systems, it is challenged to describe accurately quasi-degenerate electronic structures exhibiting strong correlations.~\cite{Cohen2012}

To overcome the limitations  of the standard DFT level of theory, various distinct methods have been proposed.~\cite{Miechlich1997,Grafenstein1998,Grimme1999,Wu1999,Zhao2005,Ukai2007,Fromager_2007,Ying2012,Roemelt2013,LiManni2014} 
Nevertheless, the common aspect of all these post-DFT approaches is the application of a numerical method which is dedicated to handle the static correlations of the problem  at some point of the theory.
Besides configuration interaction based methods and variants of the complete active space self-consistent field approach, e.g. Refs.~[\onlinecite{Roos1987,Shepard1987, Lowdin1999, Werner2007, Shamasundar2011, Szalay2012,Helgaker2014, Lischka2018}],  density matrix renormalization group (DMRG) procedure is an ideal candidate for treating strong correlations.

DMRG was originally developed to describe one-dimensional quantum models in solid state pyhsics with local interactions.~\cite{White-1992b,White-1993}
The method was later generalized to treat long-range interactions found in momentum space representation of lattice models~\cite{Xiang-1996,Nishimoto-2002,Legeza-2003b} and in ab initio quantum chemistry.~\cite{White-1999,Chan-2002a,Legeza-2003a} 
The success of these developments relies on the efficient factorization of interactions~\cite{Xiang-1996,White-1999} and the optimization of the DMRG network topologies based on concepts of quantum information theory~\cite{Legeza-2003b,Rissler2006,Barcza-2011,Szalay-2015a}  
leading to tremendous reduction in computational costs.
The underlying mathematical framework, however, is not restricted to models studied in condensed matter
physics~\cite{Schollwock2005,Schollwock-2011} or applications to molecular clusters~\cite{Szalay-2015a,Olivares-2015,Baiardi2020} but among many others it can be also used to study nuclear shell models,~\cite{Dukelsky-2002,Papenbrock2005,Rotureau2006,Legeza-2015}, particles in confined potential~\cite{Legeza-2018a,Shapir-2019,Moca-2019} or problems  in the relativistic domain.~\cite{Knecht2014,Battaglia2018,Brandejs2020}
Therefore, mapping a physical problem to a proper model together with an optimal choice of basis could pave the road for DMRG applications in a broad range of disciplines which could surpass  conventional methods.

In this work, we present the application of DMRG in a  novel direction and use it as a post-DFT approach governed by the complete active space (CAS) protocol.~\cite{Cramer2005,Jensen2006} 
The discussed approach focuses on the accurate description of the static correlations treating the rest of the electronic structure on effective one-electron level.
The method, denoted as DFT-CAS-DMRG in the following, has been applied on  top of atomic KS orbitals.~\cite{Sharma2014,Takashi2017}
Here, we apply the theoretical framework on top of both atomic and periodic KS orbitals to predict the main features of the low-lying many-body spectrum of molecular and periodic ab initio systems.
In fact, quite recently, we have applied this post-DFT approach to the negatively charged boron vacancy (VB) in single layer hBN~\cite{Ivady-2019} confirming the experimental expectations on decay routes.~\cite{DyakonovArXiv2019}

As prequel of our former study, now we turn our focus to the technical aspects of this post-DFT approach adapted both to KS atomic and periodic orbitals.  
Particularly, we investigate  the vertical many-body energy spectrum of a series of finite  VB-hBN samples terminated with hydrogen atoms.
As part of the study, to the best of our  knowledge, we benchmark ab initio DMRG results on  top of plane-wave based Kohn-Sham orbitals against the vertical spectrum computed in localized basis for the first time. 
Furthermore, tendencies observed on flakes are set against the results obtained for explicit periodic sheet.

The paper is structured as follows.
In section~\ref{methods}, we introduce shortly the DMRG method and  also discuss the technical details of the  Hamiltonian matrix construction and  of the active space selection.
In section~\ref{hBN}, we describe the main features of the studied molecular systems.
In section~\ref{tech}, the numerical aspects of the simulations are summarized.
In section~\ref{results}, the results are discussed.
Finally, conclusion is provided in section~\ref{conclusions}.

\section{Methods} 
\label{methods}
\subsection{Quantum chemical Hamiltonian and molecular orbitals}
 In this study, the molecular system of $N$ electrons with coordinate $\{\rr_i\}$  is investigated  in the presence of  nuclei of atomic number $\{Z_\mu \}$ with fixed position $\{\d_\mu\}$, i.e., in the Born-Oppenheimer approximation.
 Considering the Coulomb interaction between charged particles,  the corresponding time-independent Schr\"odinger equation, 
\begin{widetext}
\begin{equation}
\hat{H}\Psi=\left[-\sum_i^N \left(  \frac{1}{2} \nabla_i ^2 
+ \sum_{\mu} \frac{Z_\mu}{\mid \rr_i-\d_\mu\mid } \right)+  \sum_{i<j}^N \frac{1}{\mid \rr_i-\rr_j \mid} + \sum_{\mu<\nu} \frac{Z_\mu Z_\nu}{\mid \d_\mu-\d_\nu \mid } \right] \Psi(\rr_1,\dots,\rr_N) =E\Psi(\rr_1,\dots,\rr_N),
\label{Ham_def}
\end{equation}
\end{widetext}
is to be solved to obtain stationary electronic wave function $\Psi(\rr_1,\dots,\rr_N)$ with energy $E$. 
Hamiltonian~\eqref{Ham_def} measured in Hartree atomic units describes the kinetic energy of the electrons, their potential energy induced by the charged nuclei in addition to the electron-electron Coulomb interaction.
Owing to the two-body interactions, the  many-electron equation, Eq.~\eqref{Ham_def}, is not separable to a system of independent single-electron equations. 
An approximate mean field level solution can be provided by the variational Hartree-Fock method which treats Hartree and exchange interactions exactly  but completely neglects correlation effects.
Based on the Hartree-Fock solution, various methods were developed to recover the correlation energy   which methods may show limitations on accessible system size or accuracy.~\cite{aszabo82_qchem}
The density functional theory (DFT) can provide a viable compromise by mapping the original  problem 
 of interacting electrons to a gas  of non-interacting particles  in an effective potential which also includes the relevant effects of the electron-electron interactions in addition to the external potentials.
 In standard DFT procedure the energy is minimized with respect to the electron density obtained iteratively from the Kohn-Sham orbitals, $\{ \psi_{ j }(\rr) \}$. 
Whereas the  Hartree potential is treated exactly in the DFT method, the accurate description of the many-orbital interactions, encoded in the exchange-correlation energy functional, is the actual challenge of the approach.
Various approximations for the functional had been suggested that were optimized in given condensed matter systems but the method became only a standard in computational chemistry with the advent of hybrid functionals which include a portion of the  exact exchange energy of the Hartree-Fock theory to give a more precise characterization of exchange effects. 

The method has numerous advantages and became extremely popular for treating weakly correlated systems in the last decades.
Notwithstanding the enormous success and broad applicability of the approach,  one may have to go beyond the pure DFT approach to provide an accurate description of strongly-correlated electronic states.
In the following, we summarize the simplest such possible procedure applicable on one-electron level theory based on the complete active space method.

\subsection{Matrix elements of the ab-initio Hamiltonian}
Considering large supercells, the electronic bands become flat, i.e., dispersionless in the reduced Brillouin zone.
Furthermore,  band states of $\Gamma$-point, which is the center of the Brillouin zone with zero momentum, are purely real-valued.
Consequently, $\Gamma$-point states of large supercells can be regarded as molecular orbitals of solid state systems. 
Hence, similarly to molecular orbitals obtained by atomic codes, electronic  states computed by periodic programs can  also be employed to construct the corresponding ab initio Hamiltonian.
The following discussion summarizes the computation of the Hamiltonian's matrix elements of spin restricted molecular orbitals and $\Gamma$-point only periodic states. 
In the following, these localized $\Gamma$-point periodic states are also referred to as orbitals.
For the sake of simplicity, spin-restricted orbital set is considered but the concept can be readily generalized.

In terms of  operators $\hat{a}^\dagger_{i\sigma}$ and $\hat{a}^{\phantom\dagger}_{i\sigma}$, which creates and annihilates electron with $\sigma$ spin projection in orbital $i$,  the second quantized representation of the  ab initio Hamiltonian operator, Eq.~\eqref{Ham_def}, is parameterized as
\begin{equation}
\hat{H}=\sum_{ij,\sigma} t^{\phantom\dagger}_{ij}\hat{a}^\dagger_{i\sigma}\hat{a}^{\phantom\dagger}_{j\sigma} + \frac{1}{2} \sum_{ijkl, \sigma\sigma'}V^{\phantom\dagger}_{ijkl} \hat{a}^\dagger_{i\sigma}\hat{a}^\dagger_{j\sigma'}\hat{a}^{\phantom\dagger}_{k\sigma'}\hat{a}^{\phantom\dagger}_{l\sigma}+E^{\rm nuc}
\label{Ham}
\end{equation}
with system specific one- and two-electron integrals, $t^{\phantom\dagger}_{ij}$, and $V_{ijkl}^{\phantom\dagger}$, respectively.
Assuming all-electron description, the former integrals expand the kinetic contributions and the   electron-nuclear attractions in the basis of   electronic orbitals, $\{ \psi_{ j }(\rr) \}$,   
\begin{equation}
t_{ij}= - \int \psi^*_{i} (\rr) \left(  \frac{1}{2}  \nabla ^2 
+ \sum_{\mu} \frac{Z_\mu}{\mid \rr-\d_\mu\mid } \right) \psi_{ j }(\rr) d\rr
 \end{equation}      
with nuclei of atomic number $Z_\mu$ at fixed coordinate $\d_\mu$. 
Note that applying pseudopotentials, which is optional in atomic codes, but rather mandatory in plane-wave based calculations, the nuclei and the frozen core electrons are described by non-local potential functions.
When states either in the band gap or close to it are considered, the correlation effects with core electron are negligible and the pseudopotential approximation is  adequate.
As a consequence of the approach, i.e., by reducing the number of electrons and making the core potential softer, plane-wave calculations become considerably less demanding. 
 
The  interelectronic repulsions yield two-electron integrals, 
\begin{equation}    
V_{ijkl}^{\phantom\dagger}=  \int \frac{\psi^*_{i} (\rr) \psi^*_{j } (\rr') 
\psi_{k } (\rr') \psi_{l } (\rr)}{\mid \rr-\rr' \mid} d\rr d\rr',
\label{two}
\end{equation}
whereas the nuclear-nuclear repulsion gives a constant energy shift,
\begin{equation}    
E^{\rm nuc}=
 \sum_{\mu<\nu} \frac{Z_\mu Z_\nu}{\mid \d_\mu-\d_\nu \mid }.
\end{equation}

\subsubsection*{Evaluation of the two-electron matrix elements}
The definition of the two-electron integrals \eqref{two} prescribes an expensive six dimensional real space integral. 
It is easy to show that the matrix elements can be equivalently evaluated by a three dimensional integral in reciprocal space $\q$,
\begin{equation}    
V_{ijkl}=  \int \frac{\rho_{il}(\q)\rho_{jk}(-\q)}{|\q|^2} d\q 
\label{two_ms}
\end{equation}
using auxiliary density operators,
\begin{equation}    
\rho_{il}(\q)=  \int \psi^*_{i} (\rr) \psi_{l } (\rr){\rm e}^{{\rm i}\q\rr} d\rr \,.
\end{equation}
We implemented the computation of Hamiltonian matrix elements within the plane-wave based Quantum Espresso suite~\cite{Giannozzi2009,QE-2017}
adopting the momentum space representation, Eq.~\eqref{two_ms}, where the evaluation reduces to two wave function multiplications and Fourier transformation besides a threefold integration.
Restricting ourselves to the usage of norm-conserving wave functions, we do not  face the  problem of augmentation charges observed in systems with ultrasoft pseudopotentials.~\cite{Kresse1994}
The integrable divergence for $|\q| \rightarrow 0$ can be analytically treated using Gygi's approach.~\cite{Gygi-1986}

The number of two-electron matrix elements to be evaluated scales as quartic of the number of active orbitals thus their effective computation is the most critical part of  constructing the Hamiltonian matrix.
In order to optimize performance, balancing between memory size and numerical efforts,  portions of the computationally expensive auxiliary operators, $\{ \rho_{il}(\q) \}$, are cached during the two-electron matrix evaluation.

\subsection{DMRG}
In the following, the basic concept of the DMRG method is summarized, whereas interested reader finds thorough overviews of DMRG and related numerical approaches in the context quantum chemistry (see Refs.~[\onlinecite{Szalay-2015a,Olivares-2015}]).

Many-body  wave function $\ket{\Psi}$, i.e., an  eigenstate of Eq.\eqref{Ham}, expanded  in the space of $L$ spatial-orbitals reads as
\begin{equation}
\ket{\Psi} = \sum\limits_{\{n\}} C^{(n)} 
 \prod_{i=1}^L\left( \hat{a}_{i\uparrow}^{\dagger} \right)^{ n_{i\uparrow} } \left( \hat{a}_{i\downarrow}^{\dagger} \right)^{ n_{i\downarrow} } \ket{0}, \label{FCI}
\end{equation}
with notation $n=(n_{1\uparrow} n_{1\downarrow} n_{2\uparrow} n_{2\downarrow} n_{3\uparrow} n_{3\downarrow} ... n_{L\uparrow} n_{L\downarrow})$ where $n_{i\sigma} \in \{0,1 \}$.
The components of the state specific $C$ tensor increase exponentially with  system size $L$ scaling as $2^{2L}$.
Hence, the exact solution of problem~\eqref{Ham}  on current classical machines is generally limited to systems represented with a dozen of orbitals.

It is shown~\cite{Ostlund-1995} that tensor $C$ is factorizable to a set of numerically more manageable matrices using the equivalent matrix product state (MPS) form of it, i.e., 
\begin{equation}
C^{(n)} = \prod_{i=1}^L A_i^{(n_{i\uparrow} n_{i\downarrow})}.
\end{equation}
The dimension of the $A_i$ matrix  grows towards the center of the MPS chain as ${\rm dim}(A_i)=[4^{i-1},4^i]$ for $i\leq L/2$ and  ${\rm dim}(A_i)=[4^{L-i+1},4^{L-i}]$  otherwise (for even $L$).
Consequently,  recovering  exact MPS representation of tensor $C$, the overall computational cost of the  MPS matrices also grows exponentially just as the original tensor representation.
To overcome this limit, the DMRG approach provides an approximate description of the $C$ tensor in terms of optimized matrices, $A_i^{\rm DMRG}$, truncated to a fixed manageable bond dimension, $M$, i.e., ${\rm dim}(A_i^{\rm DMRG})\leq[M,M]$. 
Increasing $M$, the precision of the approximation is well controlled approaching variationally the exact solution.
In the DMRG protocol the MPS matrices are locally optimized and truncated by minimizing the  discarded entanglement between the left and right neighboring blocks of the MPS chain, obtained from the reduced density matrix of the block.
The algorithm iterates through the MPS chain in sequential order back and forth  until reaching convergence.
For a detailed tutorial about the DMRG approach in MPS formalism, we refer to Ref.~[\onlinecite{Schollwock-2011}].

Compared to wave functions obtained from typical quantum chemical approaches, the DMRG method parameterizes the eigenstates in terms of local variational objects instead of excitations of a reference configuration.
Consequently, DMRG can precisely describe active spaces up to 40-80 orbitals due to the implicitly polynomial scaling of computational costs for gapped non-critical systems. 
Most typically the method is applied to obtain the ground state properties but it can be used to describe not only the vertical but also the relaxed low-lying electronic excitations as well.~\cite{Barcza-2013,Hu-2015}

In the following, we compute the vertical excitation spectrum on the relaxed ground state geometry.
In the DMRG truncation procedure, the reduced density matrix of the blocks is formed of the equally weighted linear combination of all target states.

\subsection{Complete active space (CAS) method}
The investigated systems, consisting of dozens of atoms, is described by hundreds of Kohn-Sham orbitals which cannot be  directly treated by the DMRG owing to the computational costs. 
Therefore, an optimal selection of orbitals with tractable size is needed which are responsible for the strong static correlations.
The complete active space (CAS) scheme~\cite{Cramer2005,Jensen2006} classifies the set of orbitals to three categories, i.e., the so called core and virtual orbitals are frozen to the mean field level and filled with two and zero electrons, respectively.
The third class comprises of the so called active orbitals  which are populated with the rest of electrons minimizing the energy.

Accordingly,  the virtual orbitals does not play any role in the corresponding CAS Hamiltonian  whereas the core electrons affect the electrons of the active space through the Coulomb interactions, i.e., the Hamiltonian of the active space reads
\begin{eqnarray}
\hat{H}^{\rm CAS}&=& E^{\rm nuc}+E^{\rm core} +\sum_{ij,\sigma} t^{\rm CAS}_{ij}\hat{a}^\dagger_{i\sigma}\hat{a}^{\phantom\dagger}_{j\sigma} \\\nonumber 
&+&  \frac{1}{2}\sum_{ijkl, \sigma\sigma'}V^{\phantom\dagger}_{ijkl} \hat{a}^\dagger_{i\sigma}\hat{a}^\dagger_{j\sigma'}\hat{a}^{\phantom\dagger}_{k\sigma'}\hat{a}^{\phantom\dagger}_{l\sigma}
\label{Ham-CAS}
\end{eqnarray}
with the previously defined $V^{\phantom\dagger}_{ijkl}$ integrals restricted to the active orbital set.
The one-electron integrals of the CAS space,  $t^{\rm CAS}_{ij}$, describes not only the kinetic energy  of the active electrons and their  attraction to nuclei but also their interaction  with the core electrons. 
Describing the active electrons with the DMRG method, which treats exactly the electron exchange, the one-electron interactions  are written  as
\begin{equation}
 t^{\rm CAS}_{ij}= t^{\phantom\dagger}_{ij}+\frac{1}{2}\sum_c(2V^{\phantom\dagger}_{iccj}-V^{\phantom\dagger}_{icjc})
\end{equation}
to treat the Coulombic effects of the frozen electrons on the active orbitals.
Here, the summation runs only on the indices of the core orbitals.
Finally, the additional energy contribution of the core electrons is summed up in term $E^{\rm core}$, i.e, 
\begin{equation}
 E^{\rm core}= 2\sum_{c} t^{\phantom\dagger}_{cc}+\sum_{cc'}(2V^{\phantom\dagger}_{cc'c'c}-V^{\phantom\dagger}_{cc'cc'}).
\end{equation}
Including more and more orbitals in the active space, the corresponding CAS ground state energy gets lower and lower consistently, i.e., enlarging the active space class the energy of the CAS approaches  variationally the ground state energy of the original problem.

In practice, the active space is restricted to the most important orbitals with fractional occupational numbers.
Even though the method has limitations to  provide correct description of dynamical correlations using relatively small active space, it captures  static correlations with high accuracy providing valuable insight into the low-lying energy spectrum and the essential structure and symmetry properties of the corresponding electronic eigenstates.

Note also that, contrary to alternative post-DFT methods,~\cite{BockstedteNPJ2018}  the CAS Hamiltonian~\eqref{Ham-CAS} to be investigated does not include the Kohn Sham energies explicitly but only the Kohn Sham orbitals by construction.  
Also, the absolute energies of the states computed from the CAS Hamiltonian are not trivially comparable with counterparts obtained on the DFT level of theory due to the different description of the exchange and correlation effects.

\subsection{Protocols for selecting the active space orbitals}
\label{sect:cas_select} 
\begin{figure}[t!]
\includegraphics[width=0.9\columnwidth]{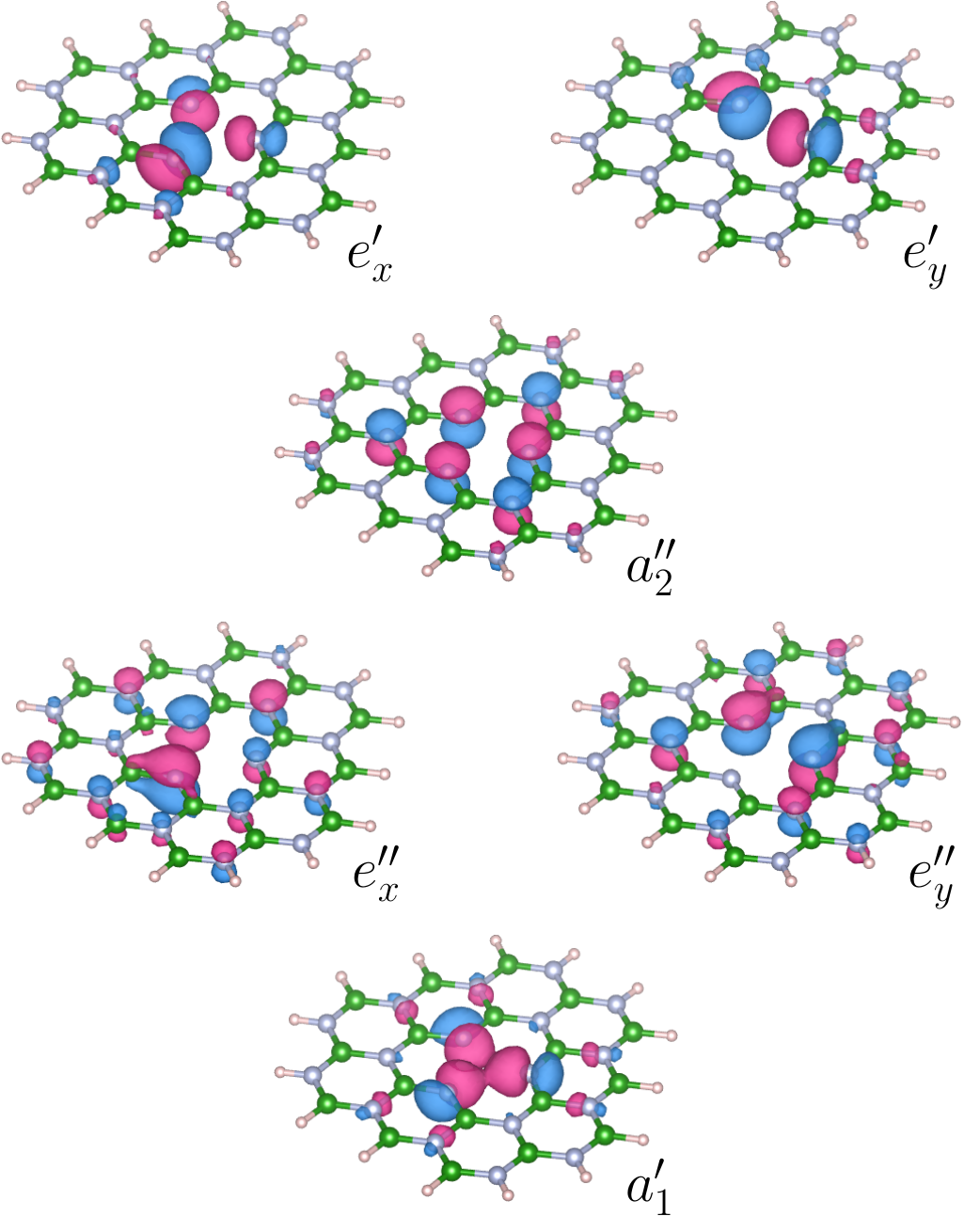}
\caption{The six  defect molecular orbitals demonstrated for  ${\rm B_{18}N_{18}H_{15}}$ in cc-pVDZ basis. Defect orbitals  characterized in Sect.~\ref{hBN} are  localized dominantly on the core of the defect, i.e., on the three neighbor  nitrogen atoms. 
\label{fig:single} 
}
\end{figure}

The studied systems are described by hundreds of orbitals, which limits the applicability of sophisticated CAS selection schemes based on large-scale post-self-consistent field calculations.~\cite{Cramer2005,Stein2016}
Nevertheless, in the studied defect systems, the low-lying many-body energy spectrum is expected to be conceptually described by the localized defect orbitals (see Fig.~\ref{fig:single}) and their interaction with the hosting environment, i.e., orbitals with large lobes around the central atoms can potentially interfere with the in-gap orbitals.
Therefore, in the following, we apply two approaches to define the active space:
i) we select canonical orbitals  with prominent localization at the core of the defect, 
ii) we construct the active space based on energy window around the valence band maximum (VBM) level to verify  the applicability of the CAS selection based on orbital localization.

It is to be noted that we do not localize the canonical orbital set in order to preserve their point group symmetry thus providing the possibility of studying the symmetry properties of the many-body states.
In practice, such orbitals  are taken into account whose degree of localization on the central part of the defect yield a critical value  according to their orbital volumetric data or (in case of availability) to their projection on atomic basis.
   
Comparing the many-body excited state spectrum yielded from the two CAS selection protocols, we find that the structure of the  spectra is  essentially identical up to some $0.001-0.2$~eV shifts in energy as discussed in section~\ref{CAS_results}, i.e., the most important orbitals are the localized in-gap  orbitals incorporated in both of the applied protocols.
Nevertheless, obtaining slightly lower absolute energies with the applied variational computational procedure indicates that the orbital selection based on localization is preferable to the one based on KS energy. 
Correspondingly, results discussed  in section~\ref{scaling_as_L} are based on DMRG calculations performed on CAS which is selected according to orbital localization.

\section{Investigated systems}
\label{hBN}
Defect-free hexagonal boron nitride monolayer has been investigated numerically on the level of ab initio dynamical mean-field theory using crystal atomic orbitals.
The computed direct and indirect band gaps are of around $10$~eV.~\cite{Zhu2020}

Various  defects embedded in hBN with potential in technology have been recently investigated, for a current review see Ref.~[\onlinecite{Sajid-2019a}].
In this work, we study hBN nano-flakes and sheet hosting a negatively charged boron vacancy (VB-hBN).
The defect system of $\rm D_{3h}$ symmetry exhibits spin triplet ground state which can be understood as following.
The in-plane dangling bonds and the out-of-plane $p_z$ orbitals of the three neighbor nitrogen atoms, located at the core of the defect, provide six single-particle defect states.
The two non-degenerate $a$ and the two degenerate $e$ defect states plotted in Fig.~\ref{fig:single} are occupied with ten electrons in the negatively charged system.
The highest-lying $e$ dangling bond becomes half occupied yielding  spin-triplet ground state.
Analysing the single-electron spectrum of VB-hBN also reveals that these localized in-gap orbitals are close in energy.~\cite{PlenioACS2018}

Therefore, the accurate description of the strongly-correlated excited states necessitates the application of multi-reference methods. 
In our recent work,~\cite{Ivady-2019} we provided a detailed description of magneto-optical properties of the system and also studied its many-body electronic spectrum applying the DFT-CAS-DMRG method.
In this work, we turn our focus on the computational aspects of the DFT-CAS-DMRG approach illustrated on VB-hBN samples.

\section{Computational details} 
\label{tech}
We study two-dimensional planar VB-hBN flakes of various sizes, i.e., investigating clusters of increasing size is necessary to capture long-range correlations and to minimize the finite-size effects of the terminating hydrogen atoms.
In particular,  we study molecules ${\rm B_{6}N_{6}H_{9}}$, ${\rm B_{18}N_{18}H_{15}}$, ${\rm B_{36}N_{36}H_{21}}$, ${\rm B_{60}N_{60}H_{27}}$, ${\rm B_{90}N_{90}H_{33}}$.
In addition, we investigate a periodic sheet of a 182 atom supercell which is large enough to minimize the interference between neighboring impurity sites and to reach convergence even restricting to $\Gamma$-point-only description. 

The hosting molecular flakes embed the vacancy in the center of the model so that the relaxed lattice structures exhibit $\rm D_{3h}$ symmetry.
The relaxed molecular geometries of the charged defects are obtained on DFT-PBE level of theory~\cite{PBE} with a homogeneous compensating background charge using VASP package.~\cite{VASP}
The finite flakes are embedded in a cubic supercell of 30~\AA, correspondingly, 30~\AA~vacuum size is applied in the perpendicular direction of the periodic sheet. 

The electronic structure of the molecules is described in terms of spin restricted Kohn-Sham DFT orbitals using PBE functional.
Self-consistent field (SCF) calculations are performed using both atomic basis based quantum chemical program suite ORCA~\cite{Neese-2012}  and plane-wave based Quantum Espresso (QE)~\cite{QE-2017} with  norm-conserving pseudo potentials.
The actual capabilities of the periodic code are taken advantage of studying the single layer.
Various atomic basis sets and plane-wave basis with increasing  cutoff are tested as discussed in the Sect.~\ref{analysis_B18}. 
\begin{figure}[t]
  \includegraphics[width=8cm]{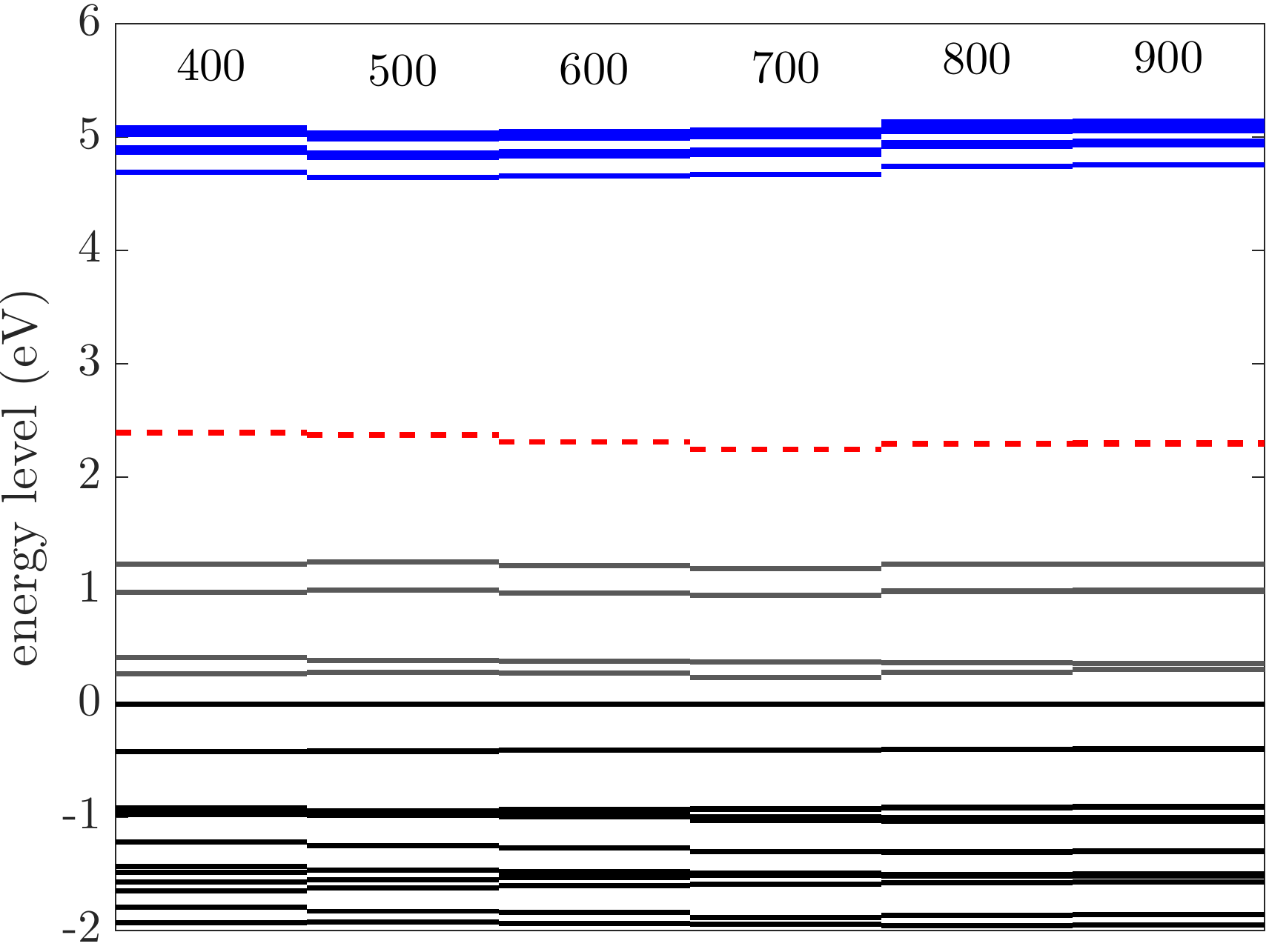}
  \caption{Snippet of KS energy spectrum close to the VBM,   for VB-hBN model with  18 boron atoms.
  DFT calculations on the DFT-PBE level of theory are performed using plane-wave basis with kinetic energy cutoff of the wave function at $\epsilon=400, 500, 600, 700, 800$ and $900$~eV. Valence, virtual, doubly/partially occupied in-gap orbitals are visualized by color black, blue, gray and red, respectively. 
 \label{fig:KSvsQE_cutoff}}
\end{figure}
\begin{figure}[t]
  \includegraphics[width=8cm]{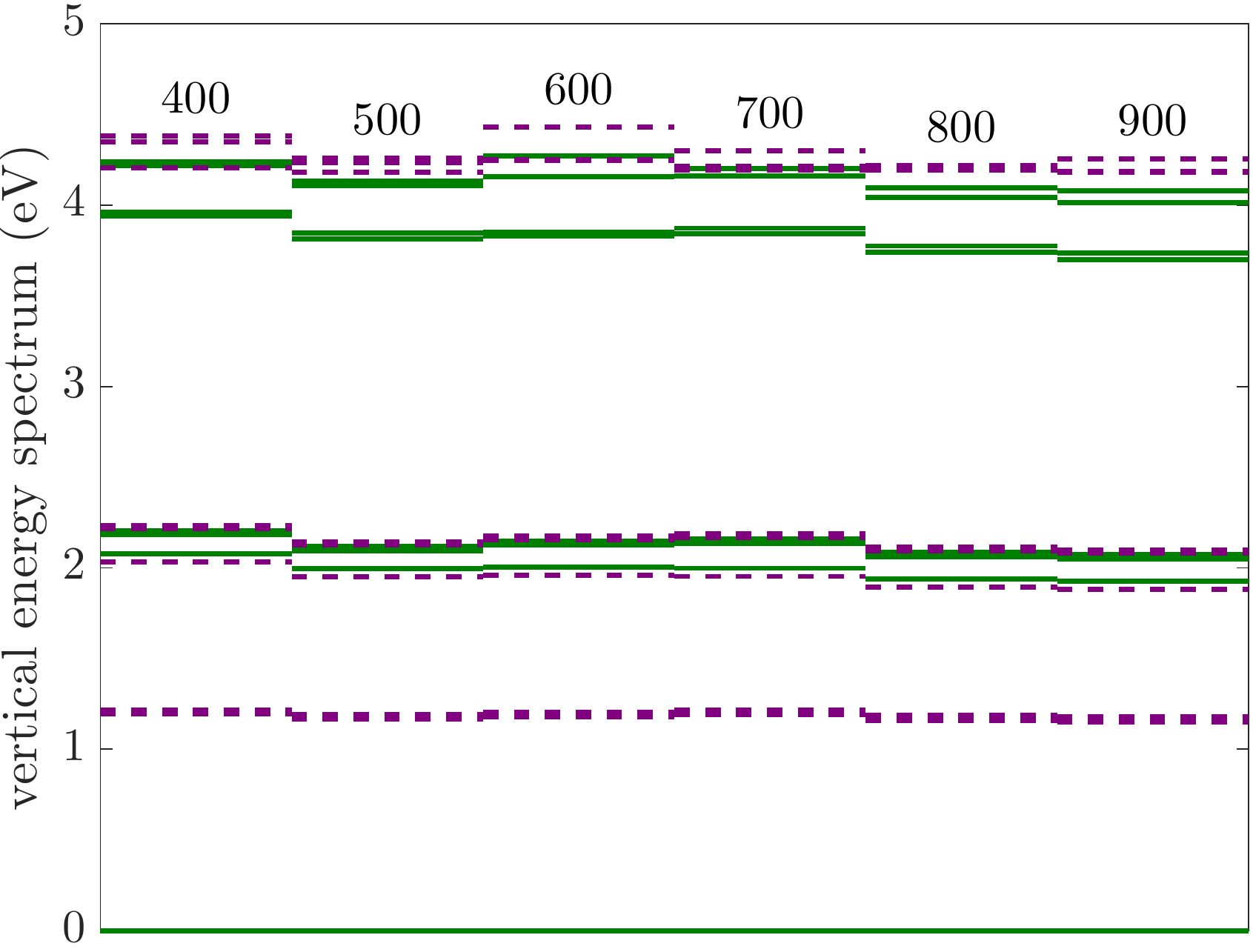}
  \caption{Vertical energy spectrum relative to the ground state energy for model ${\rm B_{18}N_{18}H_{15}}$. The active space is selected according to KS energy. Orbitals are expanded in PW basis with various energy cut-off $\epsilon$.  
   For better visibility spin-triplet and spin-singlet energy levels are contrasted using solid green and dashed purple line, respectively. 
 \label{fig:DMRGvsQE_cutoff}}
\end{figure}

The Hamiltonian matrix elements are computed by our in-house implementation for QE  KS orbitals.
Alternatively, matrix elements are also obtained from ORCA calculations through the interface to MRCC~\cite{mrcc2020,mrcc2020b} for isolated flakes. 
Active spaces of up to 40 orbitals are selected as discussed in Sect.~\ref{sect:cas_select}.

DMRG simulations are performed using the Budapest DMRG package.~\cite{budapest_qcdmrg}
Due to the a dozen of excited states to be described, DMRG calculations are performed for target states with total spin 0 and 1.
In the DMRG truncation procedure, the density matrix is constructed of the equally weighted linear combination of all target states.
As discussed in Ref.~[\onlinecite{Szalay-2015a}], the convergence is sped up by initialization procedures inspired by quantum information theoretical considerations and the accuracy of the simulations is controlled by the dynamic block state selection approach.
In the calculations, the quantum information loss is kept below threshold value $\chi=10^{-5}$. 

Note that the direct comparison with experimental data is not straightforward as the important structural relaxation effects are not incorporated in the current level of theory.
Furthermore, it is also to be kept in mind that the neglected orbital relaxation effects as well as the accurate treatment of dynamical screening effects decrease the presented vertical excitation energies.

\section{Results}
\label{results}
\subsection{Energy spectrum of \texorpdfstring{${\rm B_{18}N_{18}H_{15}}$}{} obtained in localized and in plane-wave basis sets}
\label{analysis_B18}
To demonstrate the robustness of the plane-wave based calculations, we investigate both the Kohn-Sham single-electron and the resulted vertical many-body excitation spectrum of the model with 18 boron atoms in terms of the energy cutoff, which are plotted in Figs.~\ref{fig:KSvsQE_cutoff} and \ref{fig:DMRGvsQE_cutoff}, respectively.
The kinetic energy cutoff  of the wave function is varied in the $\epsilon=400-900$~eV range whereas  the cutoff for charge density and potential is fixed as  $4\epsilon$ in agreement with the default QE parametrization.
The details of the KS spectrum around the VBM are presented in Fig.~\ref{fig:KSvsQE_cutoff} where the energy is measured relative to the lowest lying defect state.
We find that the energy spectrum is rather independent of the applied energy cut-off which also illustrates the quick convergence of the DFT calculations.
The observed small fluctuations  are acknowledged to the fact that $\epsilon$ is increased in steps of 100 and not in steps of the lowest $\epsilon=400$~eV value.
Correspondingly, due to the convergence of the KS orbitals observed already for surprisingly low cut-offs, the orbital-dependent DMRG results found in Fig.~\ref{fig:DMRGvsQE_cutoff} vary marginally for $\epsilon>400$~eV.
Therefore, in the following, according to the observed quick convergence in cutoff energy, PW results for $\epsilon=700$~eV are used as reference data.

\begin{figure}[t]
  \includegraphics[width=8cm]{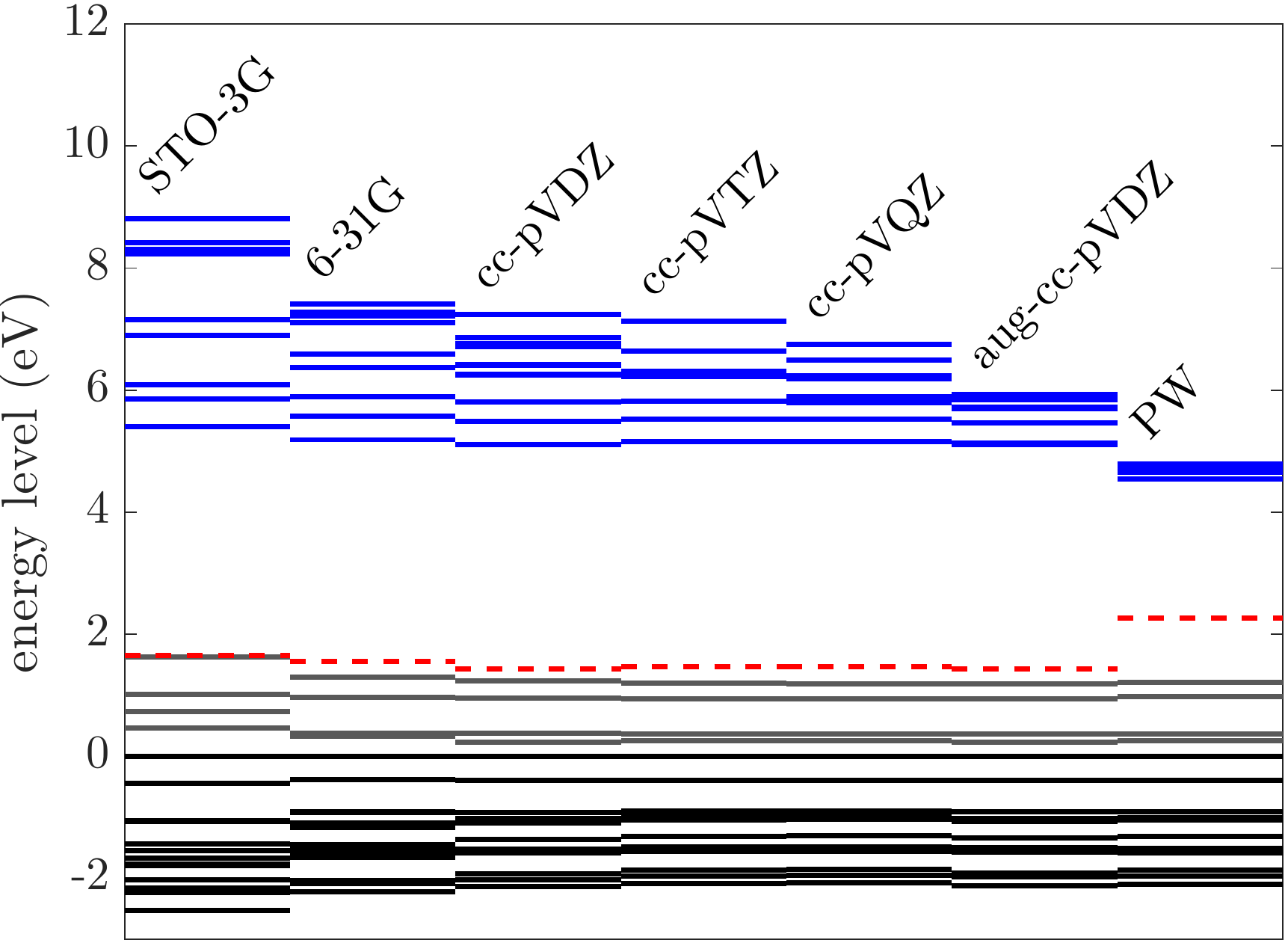} 
  \caption{Snippet of the Kohn-Sham one-electron energy spectrum of model ${\rm B_{18}N_{18}H_{15}}$ on the DFT-PBE level of theory around the gap. Orbitals are expanded in STO-3G, 6-31G, cc-pVDZ, cc-pVTZ, cc-pVQZ, aug-cc-pVDZ atomic bases and in PW basis  applying energy cutoff at $\epsilon=700$~eV.  Color key as in Fig.~\ref{fig:KSvsQE_cutoff}.
 \label{fig:ORCAKSvsbasis}}
\end{figure}

We also perform a series of DFT calculations for the model with 18 borons applying various complexity of atomic basis sets,~\cite{Cramer2005} i.e., STO-3G, 6-31G, cc-pVDZ, cc-pVTZ, cc-pVQZ,  aug-cc-pVDZ.
The corresponding Kohn-Sham one-electron spectra and, as reference, the result obtained in the plane-wave basis (PW) with energy cutoff of  $\epsilon=700$~eV are presented in  Fig.~\ref{fig:ORCAKSvsbasis}.
Interestingly, whereas the details of the spectrum show some sensitivity to the applied basis, the energy gaps of the localized state are rather independent of the chosen atomic basis for 6-31G and more complex basis.

\begin{figure}[t]
  \includegraphics[width=8cm]{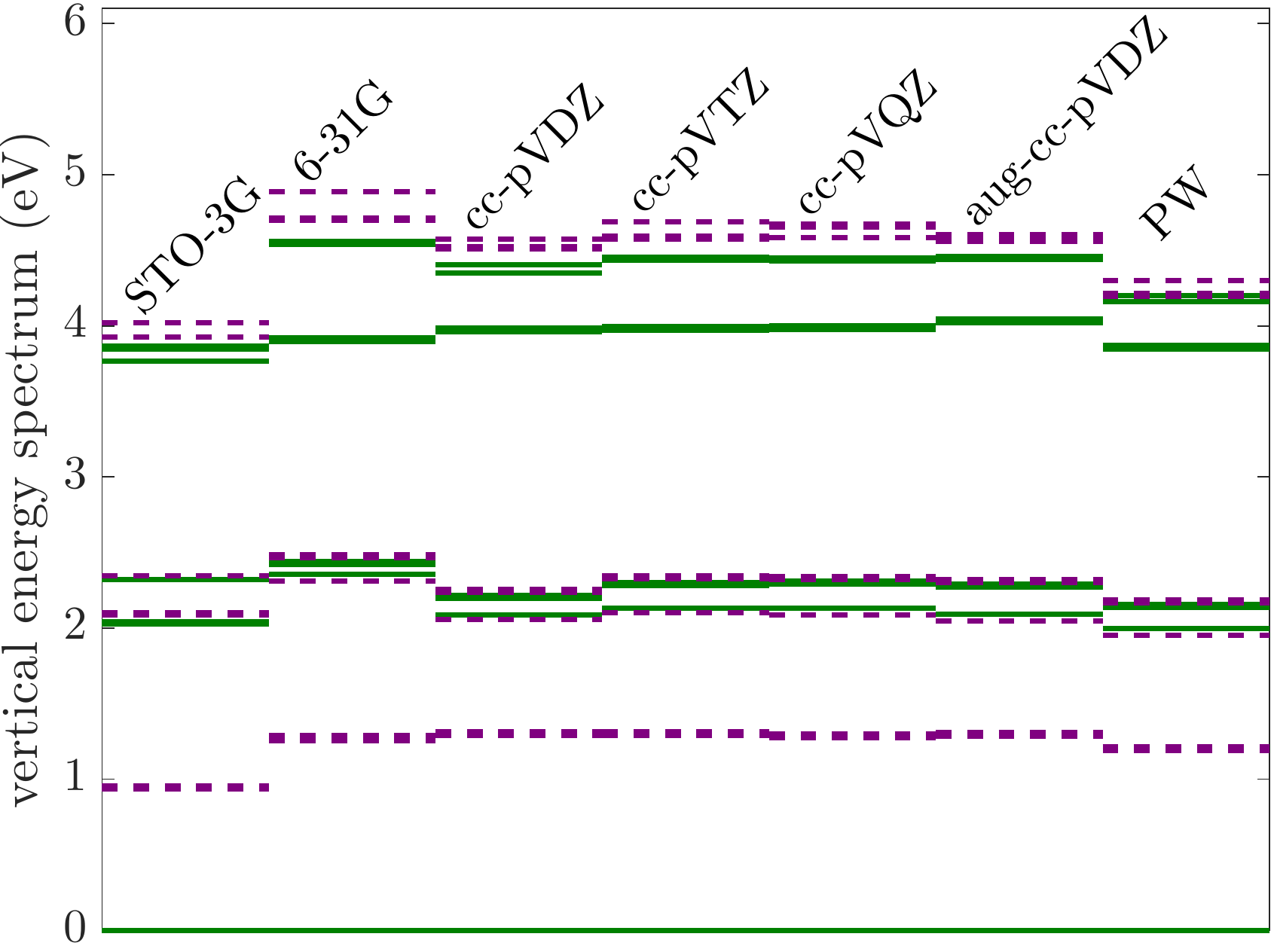}
  \caption{Relative vertical energy spectrum of model ${\rm B_{18}N_{18}H_{15}}$ obtained in active space of energy selected 21 orbitals which are expanded in STO-3G, 6-31G, cc-pVDZ, cc-pVTZ, cc-pVQZ, aug-cc-pVDZ atomic bases and also in PW basis of $\epsilon=700$~eV.   Color code as in Fig. \ref{fig:DMRGvsQE_cutoff}.
 \label{fig:DMRGvsbasis}}
\end{figure}

Comparing sophisticated atomic basis results to counterpart obtained in PW, we find great  agreement for the doubly occupied orbitals.
The observed deviations are yielded from the different treatment of the core electrons and the effect of background charge in the PW calculations. 
On the other hand, the KS energy of the partially occupied and the lowest lying virtual levels deviate significantly which understood as following. 

The energy of half-filled defect $e^{\prime\prime}$ level displayed in Fig.~\ref{fig:ORCAKSvsbasis} with dashed red line is discrepant corresponding to the distinct treatment of unpaired electrons  applied by the atomic and by the plane-wave basis computations.
In particular, the ORCA program describes the spin triplet ground state using  restricted open-shell approach whereas the unpaired electrons are set to be smeared in spin averaged manner on the degenerate $e^{\prime\prime}$ level in the QE calculation.
Therefore the observed energy difference corresponds to this spin flip energy.
Nevertheless,  we find that the corresponding ORCA and QE molecular orbitals are equivalent up to 95\%  comparing their volumetric data.
Note that such disagreement on the level of the KS spectra does not affect the presented post-DFT results as the DFT-CAS-DMRG procedure does not rely on KS energies but exclusively on orbitals. 
The discrepancy of the virtual spectra is understood as the artifact of the PW calculations where the resulted lowest lying virtuals show a tendency to describe ionization, i.e., the corresponding orbitals are localized off the molecular sheet. 
  
The many-electron DMRG energy spectrum resulted in the CAS constructed of the series of  KS-DFT-SCF orbitals is plotted in Fig.~\ref{fig:DMRGvsbasis}.
The active space for all the testing CAS calculations were selected according to the one-electron energies correlating 32 electrons in 21 orbitals.
The overall structure of the many-electron spectrum obtained in reasonably large basis sets, i.e., cc-pVDZ and beyond,  is rather  independent of the applied basis even for excitations of $4-5$~eV.
Using more and more sophisticated atomic basis, the absolute energy of each excitation lowers but with varying tendency. 
Nevertheless, approaching the completeness of the basis, i.e., using correlation consistent and PW basis with large cutoff $\epsilon$, the spectrum becomes consistent with slight variations yielding $\sim$90\% agreement on the many-body excitation energies.
It is also notable that the singlet-triplet gap, which involves predominantly spin excitation of the partially filled  defect $e^{\prime\prime}$ level, is predicted surprisingly well, i.e., with 10\% accuracy, even on the minimal STO-3G level of theory.
Based on these results, cc-pVDZ  atomic basis is an optimal choice for reasonably accurate non-periodic DFT-CAS-DMRG calculations. 

The observed discrepancies in the spectrum obtained for various orbital sets reflect not only the distinct character of the underlying incomplete basis sets but also the limitations of the CAS based description.
In particular, as valence orbitals can be very close in energy with accidental ordering as observed in Fig.~\ref{fig:ORCAKSvsbasis}, the  selection of a dozen of valence orbitals for the CAS can be biased which could be cured with the study of a much larger and optimally constructed active space.

\subsection{Sensitivity of active space selection}
\label{CAS_results}
Having a CAS based description in our hands, careful selection of active orbitals is crucial.
In the following CAS selection is investigated from the perspective of active space size and selection protocols on the example of ${\rm B_{36}N_{36}H_{21}}$ on cc-pVDZ level of theory.

First, the large gap, $\Delta\approx 5$~eV, also observed in Fig.~\ref{fig:ORCAKSvsbasis} and the partial occupation of the defect $e^{\prime\prime}$ level suggest that virtual orbitals does not play crucial role in the description of low-lying electronic states excited with energy of a few eV.  
Similar conclusion is drawn from the molecular orbitals obtained in PW SCF calculations, i.e., some of the lowest-lying virtuals are out of the plane of the sheet and expand in the vacuum, i.e., the system can be ionized with around energy $\Delta$. 

\begin{figure}[t]
  \includegraphics[width=8cm]{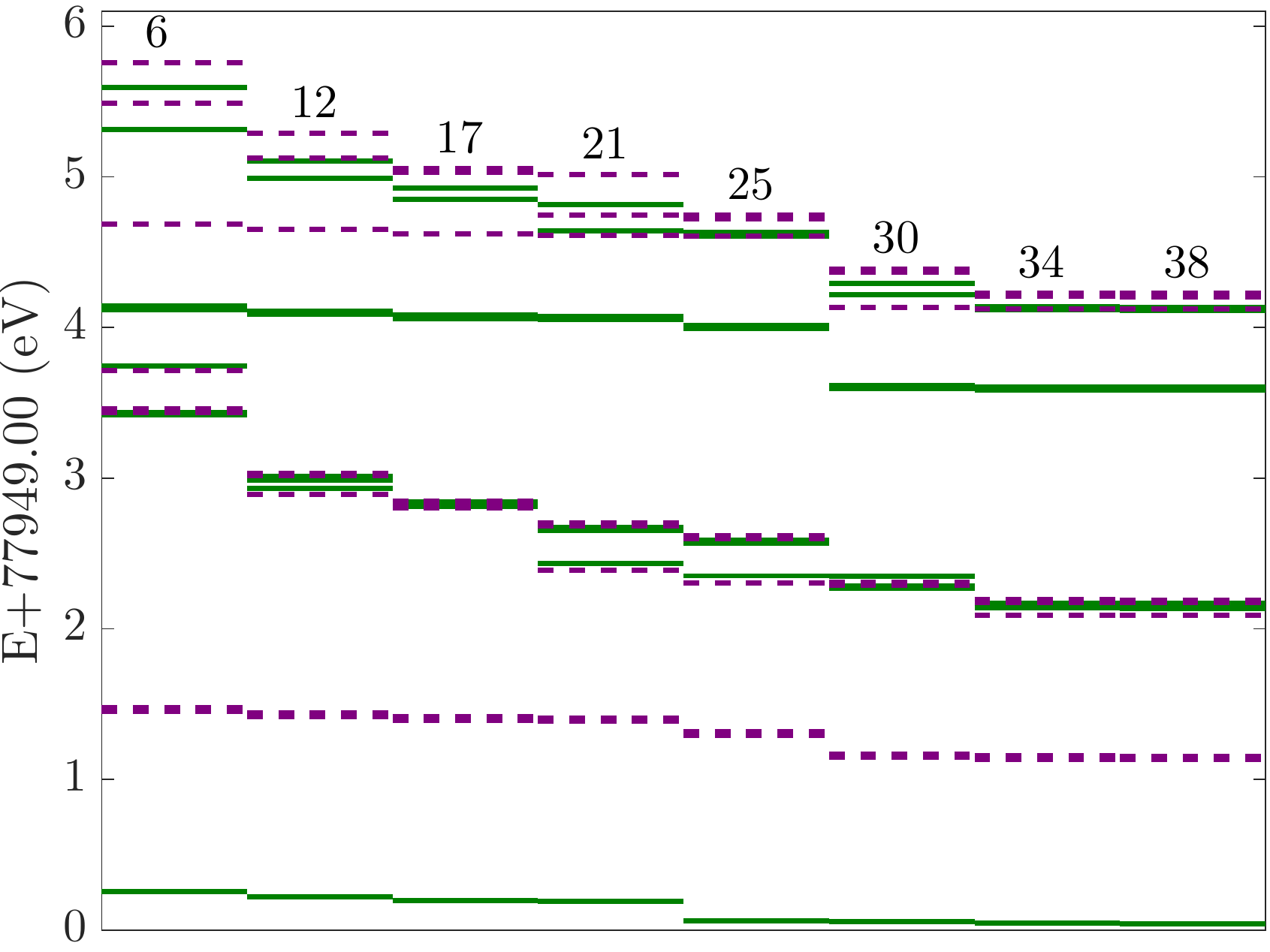}
  \caption{Many-electron energy spectrum of model ${\rm B_{36}N_{36}H_{21}}$ obtained for CAS with  6, 12, 17, 21, 25, 30, 34  and  38 orbitals selected according to KS energy on DFT-cc-pVDZ-PBE level of theory. Energies are shifted for better visibility.
    Color code as in Fig. \ref{fig:DMRGvsQE_cutoff}.
 \label{fig:DMRG_CAS_vs_size}}
\end{figure}

Hence, we typically include only a couple of virtual orbitals in the active space and select active orbitals mostly from the valence band besides the six localized orbitals presented in Fig.~\ref{fig:single}.
As an illustration, we show many-electron energy spectrum obtained for KS-energy-selected active space of  
various sizes in Fig.~\ref{fig:DMRG_CAS_vs_size}.
Increasing CAS size as more and more correlation effect is possible to be retrieved, consequently every many-electron energy decreases but with a particular rate depending on the structure of the corresponding many-body state.
We find that the main features of the spectrum are already captured using the six orbital CAS, nevertheless the energy of some many-body states lower with more than $1$~eV including more dynamical correlations by increasing CAS size.
The results also show that a CAS of around 30 orbitals is already large enough to  describe the low-lying application-relevant excitations.

It is notable that besides the ground state there are some  excitations of prominent single-reference character  whose energy decreases less dramatically with increasing active space. 
In particular, the lowest-lying spin-singlet spatial-doublet state is mainly characterized by spin excitations within the degenerate defect $e^{\prime\prime}$ level as we have shown in Ref.~[\onlinecite{Ivady-2019}].
For smaller active space sizes, i.e. $<30$, the spin-triplet doublet state with single-reference character found around $4$~eV describes excitation from the localized $a_1'$, similarly to the singlet state detected around $4.5$~eV.
Nevertheless,  extending the active space further with lower-energy orbitals, which exhibit significant localization around the defect, even these states are shifted by $0.5$~eV for active spaces of $\geq30$ orbitals.

\begin{figure}[t]
  \includegraphics[width=8cm]{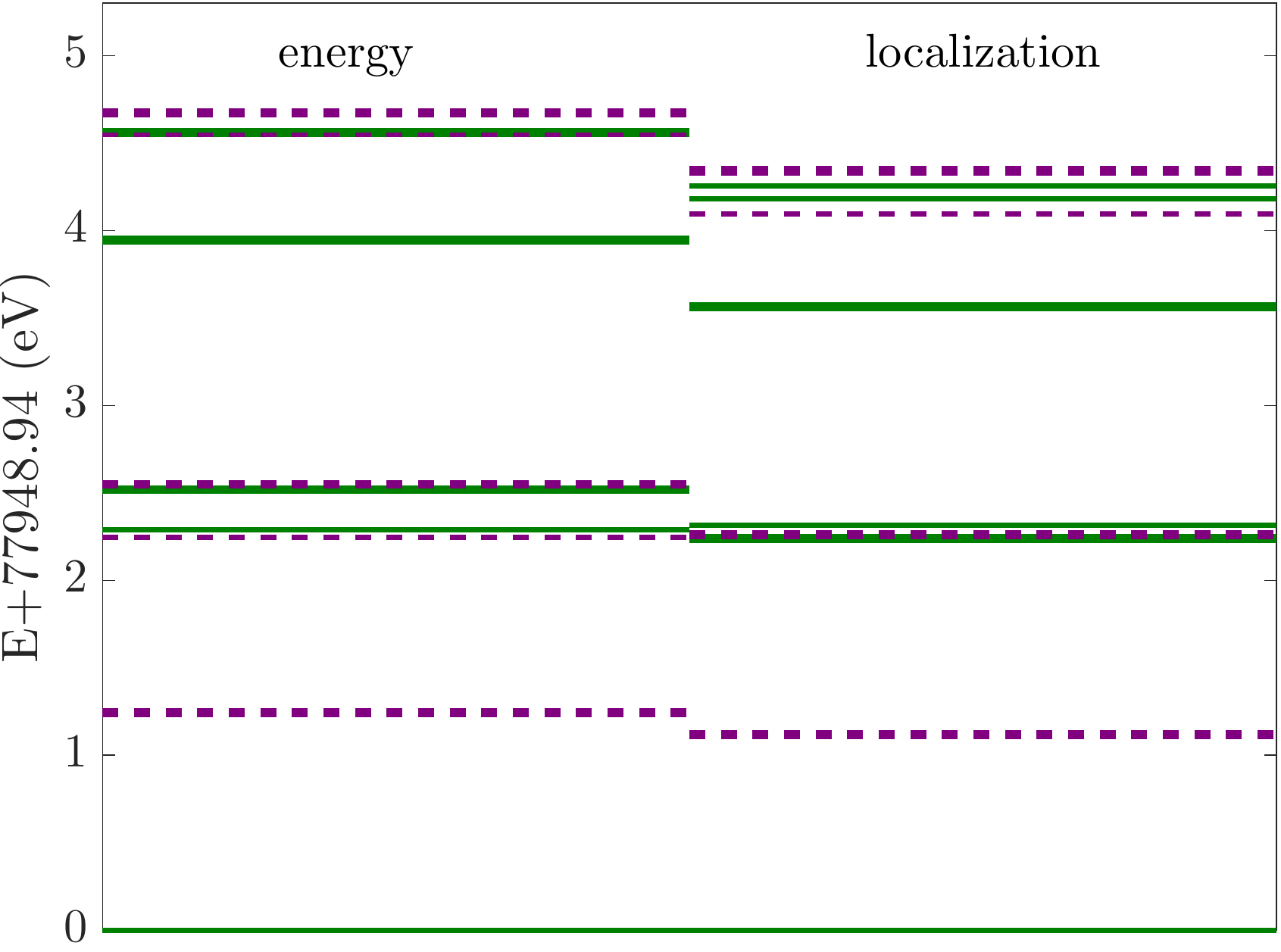}
  \caption{Many-electron energy spectrum of model ${\rm B_{36}N_{36}H_{21}}$  on DFT-cc-pVDZ-PBE level of theory  obtained for CAS of 25 orbitals based on KS energy and localization as discussed in Sect.~\ref{sect:cas_select}.
  Energies are shifted for better visibility.
   Color code as in Fig. \ref{fig:DMRGvsQE_cutoff}.
 \label{fig:DMRG_CAS_vs_method}}
\end{figure}
This observation also underlines the detailed analysis of the eigenstates, which indicates that valence orbitals showing larger overlap with the defect orbitals become more correlated in the post-DFT calculations than orbitals far from the center confirming the chemical intuition of CAS selection based on localization.
In Fig. \ref{fig:DMRG_CAS_vs_method}, we demonstrate the results obtained for 25 active orbitals selected by the two CAS construction schemes, i.e., selecting orbitals based on their KS energy and on their localization.
In the first case the occupied orbitals are kept in the active space based solely on their proximity to the Fermi surface. 
In the latter case the localization filters further the orbitals, i.e., only those orbitals are considered whose atomic basis projection  on the first neighbor three nitrogen atoms reaches at least 0.005-0.01.
Comparing DMRG spectrum obtained in the two differently constructed active spaces of 25 orbitals, corroborates the expectations, i.e. all the many-electron states get lower in energy with $0.0001-0.2$~eV using localization based CAS.
In fact, the latter calculation  recovers  essentially the spectrum obtained for the larger CAS of 30 orbitals selected according to KS energy (see Fig.~\ref{fig:DMRG_CAS_vs_size}).
In larger systems, where the corresponding orbitals are more distributed in space, the CAS selection scheme has even more drastic effect.

\subsection{Large model and single sheet results}
\label{scaling_as_L}
\begin{figure}[t]
  \includegraphics[width=8cm]{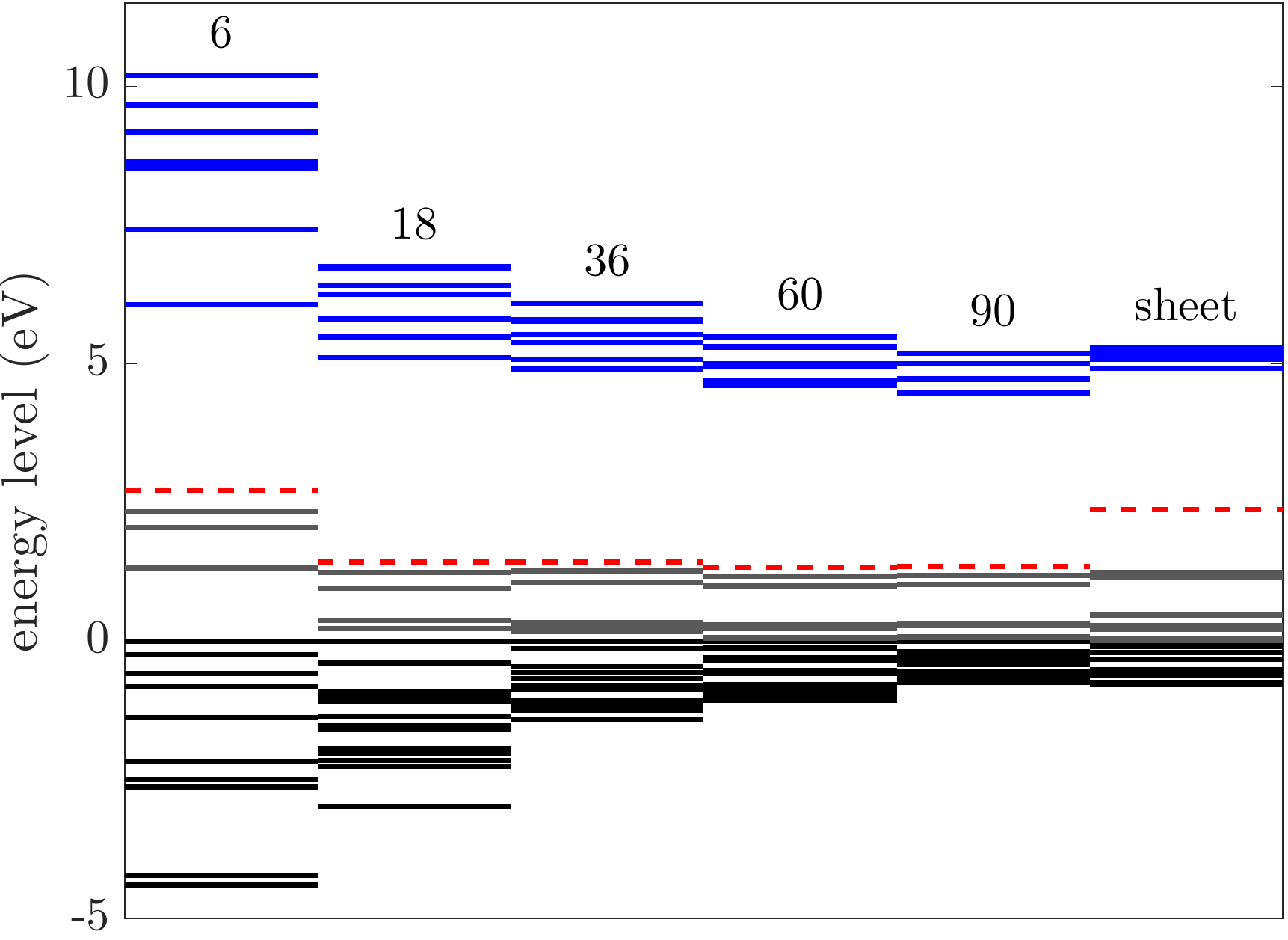}
  \caption{Snippet of the KS energy spectrum of orbitals around the gap on the DFT-PBE level of theory  for models with 6, 18, 36, 60, 90 boron atoms  and for the periodic sheet with 80 boron atoms computed using cc-pVDZ  atomic and plane-wave basis with cut-off at $700$~eV, respectively.
  Color key as in Fig. \ref{fig:KSvsQE_cutoff}.
 \label{fig:ORCAKSvsL}}
\end{figure}

The KS-PBE energy levels close to the gap are depicted in Fig.~\ref{fig:ORCAKSvsL}  for various model sizes and for  periodic sheet with 80 boron atoms.
Having observed good convergence on the relatively cheap cc-pVDZ basis set for ${\rm B_{18}N_{18}H_{15}}$ in Sect.~\ref{analysis_B18}, all molecules are treated on the  cc-pVDZ level of theory whereas the periodic sample is computed using PW basis with cut-off at $\epsilon=700$~eV.
We find that  the structure of the in-gap spectrum is similar for models with 18, 36, 60 and 90 boron atoms, however, the energy gaps are enlarged in smaller models due to the quantum confinement.
It is also clearly observed that the  confinement affects strongly the typical extended molecular orbitals but hardly the six localized defect states which are coded essentially with color red and gray in Fig.~\ref{fig:ORCAKSvsL}.
Furthermore, we also find that the doubly occupied energy spectrum of the larger molecules is in good agreement with result obtained on the periodic sheet (also noting  differences of partial-filled and virtual energies owing to technical reasons discussed in Sect.~\ref{analysis_B18}.

\begin{figure}[t]
    \includegraphics[width=8cm]{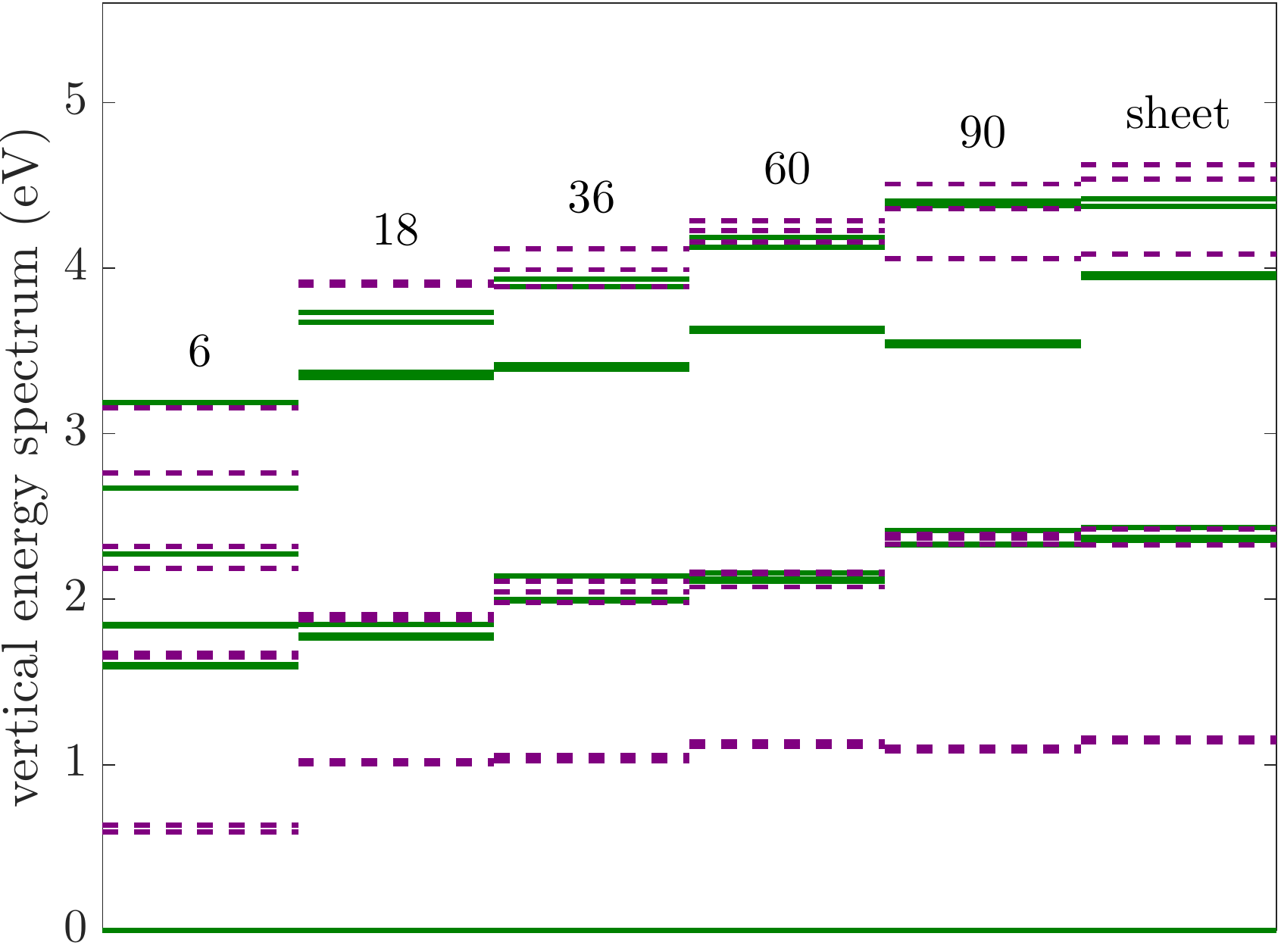}
  \caption{Relative vertical energy spectrum of models with 6, 18, 36, 60, 90 boron atoms and of  single layer with 80 borons obtained in localization-selected CAS of 40 orbitals.
  The models and the sheet are described in cc-pVDZ and PW basis, respectively.
   Color code as in Fig. \ref{fig:DMRGvsQE_cutoff}. 
 \label{fig:DMRGvsL_en}}
\end{figure}

The yielded vertical many-electron DMRG spectrum obtained for CAS with 40 localized orbitals in the active space   is presented in Fig.~\ref{fig:DMRGvsL_en} for the distinct systems.
The ground state has single reference nature in each cases, i.e. $>0.95\%$ weight for the leading configuration with half-filled $e$ orbitals is observed, confirming the applicability of DFT as ground-state calculation on VB-hBN systems.
Disregarding the smallest model with 6 borons, we find that the overall structure of the vertical many-body spectrum is largely independent of system size.
Nevertheless, the actual excitation energies generally increase  with system size yielding convergence for model with 60-90 boron atoms for lower-lying excitations.
Interestingly, the  minimal excitation energy, i.e., singlet-triplet energy gap, is less dependent on the model size.
Its stability  is the consequence of the corresponding many-body state, which is dominated by the spin flip on the half-filled twofold degenerate $e$ orbitals.

Furthermore, comparing spectra obtained on the periodic sheet with 80 boron atoms and larger flake models in Fig.~\ref{fig:DMRGvsL_en}, rather good agreement is observed. 
In particular, for excitations below $3$~eV, which are actually the most relevant  for possible quantum bit applications,~\cite{DyakonovArXiv2019} the agreement is quantitative.
Therefore, the results indicate that clusters with 60-90 borons can already reproduce the low-energy features of large periodic sheets, whereas the smaller flakes, especially ${\rm B_{6}N_{6}H_{9}}$, have limitations to capture the physics of the single layer due to the non-negligible perturbative effect of the terminating hydrogen atoms.

The effect of the active space selection on the electronic spectrum, i.e., the relevance of dynamical correlation is investigated as well.
First, in addition to the presented 40 orbital calculations, we performed further DMRG computations on 25 orbitals selected according to localization concluding very similar many-body energy spectrum. 
In fact, for models with 6-18 borons the variation is insignificant for all many-body states whereas for larger systems the energies shift with up to 5\% for the higher excitations.
Investigating the spectrum resulted of $10-15$ relevant orbitals, more significant deviations of $10-15\%$ are observed in the excitation regime above $3$~eV for larger systems. 
These results suggest that  active space of 30-40 orbitals might be already sufficient to provide reasonable description for the low-energy excitations. 
Nevertheless, the accurate description of the missed portion of dynamical correlations could potentially lower the gaps of larger systems to some extent.

\section{Conclusion}
\label{conclusions}
We report the first detailed implementation and discussion of DFT-CAS-DMRG method on plane-wave basis.
As application of the approach, we investigate the electronic structure of VB-hBN defect systems  of current interest, i.e.,  the sheet with a single defect and its molecular models.
Applying our novel computational scheme to construct the CAS Hamiltonian of QE-DFT orbitals, we find great agreement on the many-electron spectrum with results obtained using standard atomic quantum chemical program suite.
We show that flakes with around 60-90 boron atoms are large enough to reflect faithfully the low-lying energy spectrum  of the sheet with 80 borons.
In  agreement with expectations, we confirm that static correlation effects can be attributed mainly  to the six localized defect orbitals and using CAS of 30-40 orbitals can provide a reasonable description of the main electronic features of the system.
The simulation of the periodic sheet also serves as validation and benchmark of the PW based approach whose actual potential advantage  is to be realized in bulk calculations.
Given the underlying theoretical background of the presented DFT-CAS-DMRG method, we hypothesize that it might be well-suited for predicting the main features of the low-lying many-body spectrum of wide-band-gap semi-conducting defect materials with localized ingap orbitals which class of materials show emerging potentials in various quantum technology applications.

\section*{Acknowledgements}
GB acknowledges the support from the NKFIH PD-17-125261 and the Bolyai Research Scholarship of HAS. 
VI  is supported by the MTA Premium Postdoctoral Research Program and by the Knut and Alice Wallenberg Foundation through WBSQD2 project (Grant No. 2018.0071).
LV acknowledges the support from Czech Science Foundation (grant no. 18-18940Y).
\"OL and AG acknowledge support of the NKFIH through the National Quantum Technology Program (Grant No. 2017-1.2.1-NKP-2017-00001). 
\"OL acknowledges grant NKFIH K120569 and the support of the Alexander von Humboldt Foundation. 
AG  acknowledges the Hungarian NKFIH grants No. KKP129866 of the National Excellence Program of Quantum-coherent materials project. 
The computations  were performed on resources provided by the Wigner RCP and by the Swedish
National Infrastructure for Computing (SNIC 2018/3-625 and SNIC 2019/1-11) at the National
Supercomputer Centre (NSC).


\bibliographystyle{apsrev4-1}

\begin{thebibliography}{90}%
\makeatletter
\providecommand \@ifxundefined [1]{%
 \@ifx{#1\undefined}
}%
\providecommand \@ifnum [1]{%
 \ifnum #1\expandafter \@firstoftwo
 \else \expandafter \@secondoftwo
 \fi
}%
\providecommand \@ifx [1]{%
 \ifx #1\expandafter \@firstoftwo
 \else \expandafter \@secondoftwo
 \fi
}%
\providecommand \natexlab [1]{#1}%
\providecommand \enquote  [1]{``#1''}%
\providecommand \bibnamefont  [1]{#1}%
\providecommand \bibfnamefont [1]{#1}%
\providecommand \citenamefont [1]{#1}%
\providecommand \href@noop [0]{\@secondoftwo}%
\providecommand \href [0]{\begingroup \@sanitize@url \@href}%
\providecommand \@href[1]{\@@startlink{#1}\@@href}%
\providecommand \@@href[1]{\endgroup#1\@@endlink}%
\providecommand \@sanitize@url [0]{\catcode `\\12\catcode `\$12\catcode
  `\&12\catcode `\#12\catcode `\^12\catcode `\_12\catcode `\%12\relax}%
\providecommand \@@startlink[1]{}%
\providecommand \@@endlink[0]{}%
\providecommand \url  [0]{\begingroup\@sanitize@url \@url }%
\providecommand \@url [1]{\endgroup\@href {#1}{\urlprefix }}%
\providecommand \urlprefix  [0]{URL }%
\providecommand \Eprint [0]{\href }%
\providecommand \doibase [0]{http://dx.doi.org/}%
\providecommand \selectlanguage [0]{\@gobble}%
\providecommand \bibinfo  [0]{\@secondoftwo}%
\providecommand \bibfield  [0]{\@secondoftwo}%
\providecommand \translation [1]{[#1]}%
\providecommand \BibitemOpen [0]{}%
\providecommand \bibitemStop [0]{}%
\providecommand \bibitemNoStop [0]{.\EOS\space}%
\providecommand \EOS [0]{\spacefactor3000\relax}%
\providecommand \BibitemShut  [1]{\csname bibitem#1\endcsname}%
\let\auto@bib@innerbib\@empty
\bibitem [{\citenamefont {Tran}\ \emph {et~al.}(2015)\citenamefont {Tran},
  \citenamefont {Bray}, \citenamefont {Ford}, \citenamefont {Toth},\ and\
  \citenamefont {Aharonovich}}]{Aharonovich2015}%
  \BibitemOpen
  \bibfield  {author} {\bibinfo {author} {\bibfnamefont {T.~T.}\ \bibnamefont
  {Tran}}, \bibinfo {author} {\bibfnamefont {K.}~\bibnamefont {Bray}}, \bibinfo
  {author} {\bibfnamefont {M.~J.}\ \bibnamefont {Ford}}, \bibinfo {author}
  {\bibfnamefont {M.}~\bibnamefont {Toth}}, \ and\ \bibinfo {author}
  {\bibfnamefont {I.}~\bibnamefont {Aharonovich}},\ }\href
  {https://doi.org/10.1038/nnano.2015.242} {\bibfield  {journal} {\bibinfo
  {journal} {Nature Nanotechnology}\ }\textbf {\bibinfo {volume} {11}},\
  \bibinfo {pages} {37} (\bibinfo {year} {2015})}\BibitemShut {NoStop}%
\bibitem [{\citenamefont {Tran}\ \emph {et~al.}(2016)\citenamefont {Tran},
  \citenamefont {Elbadawi}, \citenamefont {Totonjian}, \citenamefont {Lobo},
  \citenamefont {Grosso}, \citenamefont {Moon}, \citenamefont {Englund},
  \citenamefont {Ford}, \citenamefont {Aharonovich},\ and\ \citenamefont
  {Toth}}]{Tran-2016}%
  \BibitemOpen
  \bibfield  {author} {\bibinfo {author} {\bibfnamefont {T.~T.}\ \bibnamefont
  {Tran}}, \bibinfo {author} {\bibfnamefont {C.}~\bibnamefont {Elbadawi}},
  \bibinfo {author} {\bibfnamefont {D.}~\bibnamefont {Totonjian}}, \bibinfo
  {author} {\bibfnamefont {C.~J.}\ \bibnamefont {Lobo}}, \bibinfo {author}
  {\bibfnamefont {G.}~\bibnamefont {Grosso}}, \bibinfo {author} {\bibfnamefont
  {H.}~\bibnamefont {Moon}}, \bibinfo {author} {\bibfnamefont {D.~R.}\
  \bibnamefont {Englund}}, \bibinfo {author} {\bibfnamefont {M.~J.}\
  \bibnamefont {Ford}}, \bibinfo {author} {\bibfnamefont {I.}~\bibnamefont
  {Aharonovich}}, \ and\ \bibinfo {author} {\bibfnamefont {M.}~\bibnamefont
  {Toth}},\ }\href {\doibase 10.1021/acsnano.6b03602} {\bibfield  {journal}
  {\bibinfo  {journal} {ACS Nano}\ }\textbf {\bibinfo {volume} {10}},\ \bibinfo
  {pages} {7331} (\bibinfo {year} {2016})}\BibitemShut {NoStop}%
\bibitem [{\citenamefont {Chejanovsky}\ \emph {et~al.}(2016)\citenamefont
  {Chejanovsky}, \citenamefont {Rezai}, \citenamefont {Paolucci}, \citenamefont
  {Kim}, \citenamefont {Rendler}, \citenamefont {Rouabeh}, \citenamefont
  {F{\'a}varo~de Oliveira}, \citenamefont {Herlinger}, \citenamefont
  {Denisenko}, \citenamefont {Yang}, \citenamefont {Gerhardt}, \citenamefont
  {Finkler}, \citenamefont {Smet},\ and\ \citenamefont
  {Wrachtrup}}]{WrachtruphBN2016}%
  \BibitemOpen
  \bibfield  {author} {\bibinfo {author} {\bibfnamefont {N.}~\bibnamefont
  {Chejanovsky}}, \bibinfo {author} {\bibfnamefont {M.}~\bibnamefont {Rezai}},
  \bibinfo {author} {\bibfnamefont {F.}~\bibnamefont {Paolucci}}, \bibinfo
  {author} {\bibfnamefont {Y.}~\bibnamefont {Kim}}, \bibinfo {author}
  {\bibfnamefont {T.}~\bibnamefont {Rendler}}, \bibinfo {author} {\bibfnamefont
  {W.}~\bibnamefont {Rouabeh}}, \bibinfo {author} {\bibfnamefont
  {F.}~\bibnamefont {F{\'a}varo~de Oliveira}}, \bibinfo {author} {\bibfnamefont
  {P.}~\bibnamefont {Herlinger}}, \bibinfo {author} {\bibfnamefont
  {A.}~\bibnamefont {Denisenko}}, \bibinfo {author} {\bibfnamefont
  {S.}~\bibnamefont {Yang}}, \bibinfo {author} {\bibfnamefont {I.}~\bibnamefont
  {Gerhardt}}, \bibinfo {author} {\bibfnamefont {A.}~\bibnamefont {Finkler}},
  \bibinfo {author} {\bibfnamefont {J.~H.}\ \bibnamefont {Smet}}, \ and\
  \bibinfo {author} {\bibfnamefont {J.}~\bibnamefont {Wrachtrup}},\ }\href
  {\doibase 10.1021/acs.nanolett.6b03268} {\bibfield  {journal} {\bibinfo
  {journal} {Nano Letters}\ }\textbf {\bibinfo {volume} {16}},\ \bibinfo
  {pages} {7037} (\bibinfo {year} {2016})}\BibitemShut {NoStop}%
\bibitem [{\citenamefont {Jungwirth}\ \emph {et~al.}(2016)\citenamefont
  {Jungwirth}, \citenamefont {Calderon}, \citenamefont {Ji}, \citenamefont
  {Spencer}, \citenamefont {Flatt{\'e}},\ and\ \citenamefont
  {Fuchs}}]{FuchshBN2016}%
  \BibitemOpen
  \bibfield  {author} {\bibinfo {author} {\bibfnamefont {N.~R.}\ \bibnamefont
  {Jungwirth}}, \bibinfo {author} {\bibfnamefont {B.}~\bibnamefont {Calderon}},
  \bibinfo {author} {\bibfnamefont {Y.}~\bibnamefont {Ji}}, \bibinfo {author}
  {\bibfnamefont {M.~G.}\ \bibnamefont {Spencer}}, \bibinfo {author}
  {\bibfnamefont {M.~E.}\ \bibnamefont {Flatt{\'e}}}, \ and\ \bibinfo {author}
  {\bibfnamefont {G.~D.}\ \bibnamefont {Fuchs}},\ }\href {\doibase
  10.1021/acs.nanolett.6b01987} {\bibfield  {journal} {\bibinfo  {journal}
  {Nano Letters}\ }\textbf {\bibinfo {volume} {16}},\ \bibinfo {pages} {6052}
  (\bibinfo {year} {2016})}\BibitemShut {NoStop}%
\bibitem [{\citenamefont {Exarhos}\ \emph {et~al.}(2017)\citenamefont
  {Exarhos}, \citenamefont {Hopper}, \citenamefont {Grote}, \citenamefont
  {Alkauskas},\ and\ \citenamefont {Bassett}}]{Bassett2017}%
  \BibitemOpen
  \bibfield  {author} {\bibinfo {author} {\bibfnamefont {A.~L.}\ \bibnamefont
  {Exarhos}}, \bibinfo {author} {\bibfnamefont {D.~A.}\ \bibnamefont {Hopper}},
  \bibinfo {author} {\bibfnamefont {R.~R.}\ \bibnamefont {Grote}}, \bibinfo
  {author} {\bibfnamefont {A.}~\bibnamefont {Alkauskas}}, \ and\ \bibinfo
  {author} {\bibfnamefont {L.~C.}\ \bibnamefont {Bassett}},\ }\href {\doibase
  10.1021/acsnano.7b00665} {\bibfield  {journal} {\bibinfo  {journal} {ACS
  Nano}\ }\textbf {\bibinfo {volume} {11}},\ \bibinfo {pages} {3328} (\bibinfo
  {year} {2017})}\BibitemShut {NoStop}%
\bibitem [{\citenamefont {Proscia}\ \emph {et~al.}(2018)\citenamefont
  {Proscia}, \citenamefont {Shotan}, \citenamefont {Jayakumar}, \citenamefont
  {Reddy}, \citenamefont {Cohen}, \citenamefont {Dollar}, \citenamefont
  {Alkauskas}, \citenamefont {Doherty}, \citenamefont {Meriles},\ and\
  \citenamefont {Menon}}]{ProsciaOptica18}%
  \BibitemOpen
  \bibfield  {author} {\bibinfo {author} {\bibfnamefont {N.~V.}\ \bibnamefont
  {Proscia}}, \bibinfo {author} {\bibfnamefont {Z.}~\bibnamefont {Shotan}},
  \bibinfo {author} {\bibfnamefont {H.}~\bibnamefont {Jayakumar}}, \bibinfo
  {author} {\bibfnamefont {P.}~\bibnamefont {Reddy}}, \bibinfo {author}
  {\bibfnamefont {C.}~\bibnamefont {Cohen}}, \bibinfo {author} {\bibfnamefont
  {M.}~\bibnamefont {Dollar}}, \bibinfo {author} {\bibfnamefont
  {A.}~\bibnamefont {Alkauskas}}, \bibinfo {author} {\bibfnamefont
  {M.}~\bibnamefont {Doherty}}, \bibinfo {author} {\bibfnamefont {C.~A.}\
  \bibnamefont {Meriles}}, \ and\ \bibinfo {author} {\bibfnamefont {V.~M.}\
  \bibnamefont {Menon}},\ }\href {\doibase 10.1364/OPTICA.5.001128} {\bibfield
  {journal} {\bibinfo  {journal} {Optica}\ }\textbf {\bibinfo {volume} {5}},\
  \bibinfo {pages} {1128} (\bibinfo {year} {2018})}\BibitemShut {NoStop}%
\bibitem [{\citenamefont {Bommer}\ and\ \citenamefont
  {Becher}(2019)}]{BecherhBN2019}%
  \BibitemOpen
  \bibfield  {author} {\bibinfo {author} {\bibfnamefont {A.}~\bibnamefont
  {Bommer}}\ and\ \bibinfo {author} {\bibfnamefont {C.}~\bibnamefont
  {Becher}},\ }\href {\doibase 10.1515/nanoph-2019-0123} {\bibfield  {journal}
  {\bibinfo  {journal} {Nanophotonics}\ }\textbf {\bibinfo {volume} {0}},\
  \bibinfo {pages} {0123} (\bibinfo {year} {2019})}\BibitemShut {NoStop}%
\bibitem [{\citenamefont {Gottscholl}\ \emph {et~al.}(2020)\citenamefont
  {Gottscholl}, \citenamefont {Kianinia}, \citenamefont {Soltamov},
  \citenamefont {Orlinskii}, \citenamefont {Mamin}, \citenamefont {Bradac},
  \citenamefont {Kasper}, \citenamefont {Krambrock}, \citenamefont {Sperlich},
  \citenamefont {Toth}, \citenamefont {Aharonovich},\ and\ \citenamefont
  {Dyakonov}}]{DyakonovArXiv2019}%
  \BibitemOpen
  \bibfield  {author} {\bibinfo {author} {\bibfnamefont {A.}~\bibnamefont
  {Gottscholl}}, \bibinfo {author} {\bibfnamefont {M.}~\bibnamefont
  {Kianinia}}, \bibinfo {author} {\bibfnamefont {V.}~\bibnamefont {Soltamov}},
  \bibinfo {author} {\bibfnamefont {S.}~\bibnamefont {Orlinskii}}, \bibinfo
  {author} {\bibfnamefont {G.}~\bibnamefont {Mamin}}, \bibinfo {author}
  {\bibfnamefont {C.}~\bibnamefont {Bradac}}, \bibinfo {author} {\bibfnamefont
  {C.}~\bibnamefont {Kasper}}, \bibinfo {author} {\bibfnamefont
  {K.}~\bibnamefont {Krambrock}}, \bibinfo {author} {\bibfnamefont
  {A.}~\bibnamefont {Sperlich}}, \bibinfo {author} {\bibfnamefont
  {M.}~\bibnamefont {Toth}}, \bibinfo {author} {\bibfnamefont {I.}~\bibnamefont
  {Aharonovich}}, \ and\ \bibinfo {author} {\bibfnamefont {V.}~\bibnamefont
  {Dyakonov}},\ }\href {https://doi.org/10.1038/s41563-020-0619-6} {\bibfield
  {journal} {\bibinfo  {journal} {Nature Materials}\ }\textbf {\bibinfo
  {volume} {19}},\ \bibinfo {pages} {540} (\bibinfo {year} {2020})}\BibitemShut
  {NoStop}%
\bibitem [{\citenamefont {Chejanovsky}\ \emph {et~al.}(2019)\citenamefont
  {Chejanovsky}, \citenamefont {Mukherjee}, \citenamefont {Kim}, \citenamefont
  {Denisenko}, \citenamefont {Finkler}, \citenamefont {Taniguchi},
  \citenamefont {Watanabe}, \citenamefont {Bhaktavatsala~Rao}, \citenamefont
  {Smet},\ and\ \citenamefont {Wrachtrup}}]{WrachtrupArXiv2019}%
  \BibitemOpen
  \bibfield  {author} {\bibinfo {author} {\bibfnamefont {N.}~\bibnamefont
  {Chejanovsky}}, \bibinfo {author} {\bibfnamefont {A.}~\bibnamefont
  {Mukherjee}}, \bibinfo {author} {\bibfnamefont {Y.}~\bibnamefont {Kim}},
  \bibinfo {author} {\bibfnamefont {A.}~\bibnamefont {Denisenko}}, \bibinfo
  {author} {\bibfnamefont {A.}~\bibnamefont {Finkler}}, \bibinfo {author}
  {\bibfnamefont {T.}~\bibnamefont {Taniguchi}}, \bibinfo {author}
  {\bibfnamefont {K.}~\bibnamefont {Watanabe}}, \bibinfo {author}
  {\bibfnamefont {D.~D.}\ \bibnamefont {Bhaktavatsala~Rao}}, \bibinfo {author}
  {\bibfnamefont {J.~H.}\ \bibnamefont {Smet}}, \ and\ \bibinfo {author}
  {\bibfnamefont {J.}~\bibnamefont {Wrachtrup}},\ }\href@noop {} {\enquote
  {\bibinfo {title} {{Single spin resonance in a van der Waals embedded
  paramagnetic defect}},}\ } (\bibinfo {year} {2019}),\ \Eprint
  {http://arxiv.org/abs/arXiv:1906.05903} {arXiv:1906.05903} \BibitemShut
  {NoStop}%
\bibitem [{\citenamefont {Orellana}\ and\ \citenamefont
  {Chacham}(2001)}]{Orellana-2001}%
  \BibitemOpen
  \bibfield  {author} {\bibinfo {author} {\bibfnamefont {W.}~\bibnamefont
  {Orellana}}\ and\ \bibinfo {author} {\bibfnamefont {H.}~\bibnamefont
  {Chacham}},\ }\href {\doibase 10.1103/PhysRevB.63.125205} {\bibfield
  {journal} {\bibinfo  {journal} {Phys. Rev. B}\ }\textbf {\bibinfo {volume}
  {63}},\ \bibinfo {pages} {125205} (\bibinfo {year} {2001})}\BibitemShut
  {NoStop}%
\bibitem [{\citenamefont {Piquini}\ \emph {et~al.}(2005)\citenamefont
  {Piquini}, \citenamefont {Baierle}, \citenamefont {Schmidt},\ and\
  \citenamefont {Fazzio}}]{Piquini-2005}%
  \BibitemOpen
  \bibfield  {author} {\bibinfo {author} {\bibfnamefont {P.}~\bibnamefont
  {Piquini}}, \bibinfo {author} {\bibfnamefont {R.~J.}\ \bibnamefont
  {Baierle}}, \bibinfo {author} {\bibfnamefont {T.~M.}\ \bibnamefont
  {Schmidt}}, \ and\ \bibinfo {author} {\bibfnamefont {A.}~\bibnamefont
  {Fazzio}},\ }\href {\doibase 10.1088/0957-4484/16/6/035} {\bibfield
  {journal} {\bibinfo  {journal} {Nanotechnology}\ }\textbf {\bibinfo {volume}
  {16}},\ \bibinfo {pages} {827} (\bibinfo {year} {2005})}\BibitemShut
  {NoStop}%
\bibitem [{\citenamefont {Si}\ and\ \citenamefont {Xue}(2007)}]{Si-2007}%
  \BibitemOpen
  \bibfield  {author} {\bibinfo {author} {\bibfnamefont {M.~S.}\ \bibnamefont
  {Si}}\ and\ \bibinfo {author} {\bibfnamefont {D.~S.}\ \bibnamefont {Xue}},\
  }\href {\doibase 10.1103/PhysRevB.75.193409} {\bibfield  {journal} {\bibinfo
  {journal} {Phys. Rev. B}\ }\textbf {\bibinfo {volume} {75}},\ \bibinfo
  {pages} {193409} (\bibinfo {year} {2007})}\BibitemShut {NoStop}%
\bibitem [{\citenamefont {Attaccalite}\ \emph {et~al.}(2011)\citenamefont
  {Attaccalite}, \citenamefont {Bockstedte}, \citenamefont {Marini},
  \citenamefont {Rubio},\ and\ \citenamefont {Wirtz}}]{Bockstedte2011hBN}%
  \BibitemOpen
  \bibfield  {author} {\bibinfo {author} {\bibfnamefont {C.}~\bibnamefont
  {Attaccalite}}, \bibinfo {author} {\bibfnamefont {M.}~\bibnamefont
  {Bockstedte}}, \bibinfo {author} {\bibfnamefont {A.}~\bibnamefont {Marini}},
  \bibinfo {author} {\bibfnamefont {A.}~\bibnamefont {Rubio}}, \ and\ \bibinfo
  {author} {\bibfnamefont {L.}~\bibnamefont {Wirtz}},\ }\href {\doibase
  10.1103/PhysRevB.83.144115} {\bibfield  {journal} {\bibinfo  {journal} {Phys.
  Rev. B}\ }\textbf {\bibinfo {volume} {83}},\ \bibinfo {pages} {144115}
  (\bibinfo {year} {2011})}\BibitemShut {NoStop}%
\bibitem [{\citenamefont {Huang}\ and\ \citenamefont {Lee}(2012)}]{Huang-2012}%
  \BibitemOpen
  \bibfield  {author} {\bibinfo {author} {\bibfnamefont {B.}~\bibnamefont
  {Huang}}\ and\ \bibinfo {author} {\bibfnamefont {H.}~\bibnamefont {Lee}},\
  }\href {\doibase 10.1103/PhysRevB.86.245406} {\bibfield  {journal} {\bibinfo
  {journal} {Phys. Rev. B}\ }\textbf {\bibinfo {volume} {86}},\ \bibinfo
  {pages} {245406} (\bibinfo {year} {2012})}\BibitemShut {NoStop}%
\bibitem [{\citenamefont {Zhang}\ \emph {et~al.}(2013)\citenamefont {Zhang},
  \citenamefont {Zhou},\ and\ \citenamefont {Zuo}}]{Zhang-2013}%
  \BibitemOpen
  \bibfield  {author} {\bibinfo {author} {\bibfnamefont {Z.-F.}\ \bibnamefont
  {Zhang}}, \bibinfo {author} {\bibfnamefont {T.-G.}\ \bibnamefont {Zhou}}, \
  and\ \bibinfo {author} {\bibfnamefont {X.}~\bibnamefont {Zuo}},\ }\href
  {\doibase 10.7498/aps.62.083102} {\bibfield  {journal} {\bibinfo  {journal}
  {Acta Physica Sinica}\ }\textbf {\bibinfo {volume} {62}},\ \bibinfo {eid}
  {83102} (\bibinfo {year} {2013})}\BibitemShut {NoStop}%
\bibitem [{\citenamefont {Cheng}\ \emph {et~al.}(2017)\citenamefont {Cheng},
  \citenamefont {Zhang}, \citenamefont {Yan}, \citenamefont {Huang},
  \citenamefont {Huang}, \citenamefont {Song}, \citenamefont {Chen},\ and\
  \citenamefont {Tang}}]{Cheng-2017}%
  \BibitemOpen
  \bibfield  {author} {\bibinfo {author} {\bibfnamefont {G.}~\bibnamefont
  {Cheng}}, \bibinfo {author} {\bibfnamefont {Y.}~\bibnamefont {Zhang}},
  \bibinfo {author} {\bibfnamefont {L.}~\bibnamefont {Yan}}, \bibinfo {author}
  {\bibfnamefont {H.}~\bibnamefont {Huang}}, \bibinfo {author} {\bibfnamefont
  {Q.}~\bibnamefont {Huang}}, \bibinfo {author} {\bibfnamefont
  {Y.}~\bibnamefont {Song}}, \bibinfo {author} {\bibfnamefont {Y.}~\bibnamefont
  {Chen}}, \ and\ \bibinfo {author} {\bibfnamefont {Z.}~\bibnamefont {Tang}},\
  }\href {\doibase https://doi.org/10.1016/j.commatsci.2016.12.032} {\bibfield
  {journal} {\bibinfo  {journal} {Computational Materials Science}\ }\textbf
  {\bibinfo {volume} {129}},\ \bibinfo {pages} {247 } (\bibinfo {year}
  {2017})}\BibitemShut {NoStop}%
\bibitem [{\citenamefont {Tawfik}\ \emph {et~al.}(2017)\citenamefont {Tawfik},
  \citenamefont {Ali}, \citenamefont {Fronzi}, \citenamefont {Kianinia},
  \citenamefont {Tran}, \citenamefont {Stampfl}, \citenamefont {Aharonovich},
  \citenamefont {Toth},\ and\ \citenamefont {Ford}}]{MFord2017}%
  \BibitemOpen
  \bibfield  {author} {\bibinfo {author} {\bibfnamefont {S.~A.}\ \bibnamefont
  {Tawfik}}, \bibinfo {author} {\bibfnamefont {S.}~\bibnamefont {Ali}},
  \bibinfo {author} {\bibfnamefont {M.}~\bibnamefont {Fronzi}}, \bibinfo
  {author} {\bibfnamefont {M.}~\bibnamefont {Kianinia}}, \bibinfo {author}
  {\bibfnamefont {T.~T.}\ \bibnamefont {Tran}}, \bibinfo {author}
  {\bibfnamefont {C.}~\bibnamefont {Stampfl}}, \bibinfo {author} {\bibfnamefont
  {I.}~\bibnamefont {Aharonovich}}, \bibinfo {author} {\bibfnamefont
  {M.}~\bibnamefont {Toth}}, \ and\ \bibinfo {author} {\bibfnamefont {M.~J.}\
  \bibnamefont {Ford}},\ }\href {\doibase 10.1039/C7NR04270A} {\bibfield
  {journal} {\bibinfo  {journal} {Nanoscale}\ }\textbf {\bibinfo {volume}
  {9}},\ \bibinfo {pages} {13575} (\bibinfo {year} {2017})}\BibitemShut
  {NoStop}%
\bibitem [{\citenamefont {Weston}\ \emph {et~al.}(2018)\citenamefont {Weston},
  \citenamefont {Wickramaratne}, \citenamefont {Mackoit}, \citenamefont
  {Alkauskas},\ and\ \citenamefont {Van~de Walle}}]{WestonPhysRevB2018}%
  \BibitemOpen
  \bibfield  {author} {\bibinfo {author} {\bibfnamefont {L.}~\bibnamefont
  {Weston}}, \bibinfo {author} {\bibfnamefont {D.}~\bibnamefont
  {Wickramaratne}}, \bibinfo {author} {\bibfnamefont {M.}~\bibnamefont
  {Mackoit}}, \bibinfo {author} {\bibfnamefont {A.}~\bibnamefont {Alkauskas}},
  \ and\ \bibinfo {author} {\bibfnamefont {C.~G.}\ \bibnamefont {Van~de
  Walle}},\ }\href {\doibase 10.1103/PhysRevB.97.214104} {\bibfield  {journal}
  {\bibinfo  {journal} {Phys. Rev. B}\ }\textbf {\bibinfo {volume} {97}},\
  \bibinfo {pages} {214104} (\bibinfo {year} {2018})}\BibitemShut {NoStop}%
\bibitem [{\citenamefont {Abdi}\ \emph {et~al.}(2018)\citenamefont {Abdi},
  \citenamefont {Chou}, \citenamefont {Gali},\ and\ \citenamefont
  {Plenio}}]{PlenioACS2018}%
  \BibitemOpen
  \bibfield  {author} {\bibinfo {author} {\bibfnamefont {M.}~\bibnamefont
  {Abdi}}, \bibinfo {author} {\bibfnamefont {J.-P.}\ \bibnamefont {Chou}},
  \bibinfo {author} {\bibfnamefont {A.}~\bibnamefont {Gali}}, \ and\ \bibinfo
  {author} {\bibfnamefont {M.~B.}\ \bibnamefont {Plenio}},\ }\href {\doibase
  10.1021/acsphotonics.7b01442} {\bibfield  {journal} {\bibinfo  {journal} {ACS
  Photonics}\ }\textbf {\bibinfo {volume} {5}},\ \bibinfo {pages} {1967}
  (\bibinfo {year} {2018})}\BibitemShut {NoStop}%
\bibitem [{\citenamefont {Sajid}\ \emph {et~al.}(2018)\citenamefont {Sajid},
  \citenamefont {Reimers},\ and\ \citenamefont {Ford}}]{Sajid2018}%
  \BibitemOpen
  \bibfield  {author} {\bibinfo {author} {\bibfnamefont {A.}~\bibnamefont
  {Sajid}}, \bibinfo {author} {\bibfnamefont {J.~R.}\ \bibnamefont {Reimers}},
  \ and\ \bibinfo {author} {\bibfnamefont {M.~J.}\ \bibnamefont {Ford}},\
  }\href {\doibase 10.1103/PhysRevB.97.064101} {\bibfield  {journal} {\bibinfo
  {journal} {Phys. Rev. B}\ }\textbf {\bibinfo {volume} {97}},\ \bibinfo
  {pages} {064101} (\bibinfo {year} {2018})}\BibitemShut {NoStop}%
\bibitem [{\citenamefont {Reimers}\ \emph {et~al.}(2018)\citenamefont
  {Reimers}, \citenamefont {Sajid}, \citenamefont {Kobayashi},\ and\
  \citenamefont {Ford}}]{Reimers-2018}%
  \BibitemOpen
  \bibfield  {author} {\bibinfo {author} {\bibfnamefont {J.~R.}\ \bibnamefont
  {Reimers}}, \bibinfo {author} {\bibfnamefont {A.}~\bibnamefont {Sajid}},
  \bibinfo {author} {\bibfnamefont {R.}~\bibnamefont {Kobayashi}}, \ and\
  \bibinfo {author} {\bibfnamefont {M.~J.}\ \bibnamefont {Ford}},\ }\href
  {\doibase 10.1021/acs.jctc.7b01072} {\bibfield  {journal} {\bibinfo
  {journal} {Journal of Chemical Theory and Computation}\ }\textbf {\bibinfo
  {volume} {14}},\ \bibinfo {pages} {1602} (\bibinfo {year}
  {2018})}\BibitemShut {NoStop}%
\bibitem [{\citenamefont {Sajid}\ \emph {et~al.}(2019)\citenamefont {Sajid},
  \citenamefont {Reimers}, \citenamefont {Thygesen},\ and\ \citenamefont
  {Ford}}]{Sajid-2019b}%
  \BibitemOpen
  \bibfield  {author} {\bibinfo {author} {\bibfnamefont {A.}~\bibnamefont
  {Sajid}}, \bibinfo {author} {\bibfnamefont {J.~R.}\ \bibnamefont {Reimers}},
  \bibinfo {author} {\bibfnamefont {K.~S.}\ \bibnamefont {Thygesen}}, \ and\
  \bibinfo {author} {\bibfnamefont {M.~J.}\ \bibnamefont {Ford}},\ }\href@noop
  {} {\enquote {\bibinfo {title} {Identification of defects responsible for
  optically detected magnetic resonance in hexagonal boron nitride},}\ }
  (\bibinfo {year} {2019}),\ \Eprint {http://arxiv.org/abs/arXiv:1912.07816}
  {arXiv:1912.07816} \BibitemShut {NoStop}%
\bibitem [{\citenamefont {Kohn}\ and\ \citenamefont {Sham}(1965)}]{Kohn1965}%
  \BibitemOpen
  \bibfield  {author} {\bibinfo {author} {\bibfnamefont {W.}~\bibnamefont
  {Kohn}}\ and\ \bibinfo {author} {\bibfnamefont {L.~J.}\ \bibnamefont
  {Sham}},\ }\href {\doibase 10.1103/PhysRev.140.A1133} {\bibfield  {journal}
  {\bibinfo  {journal} {Phys. Rev.}\ }\textbf {\bibinfo {volume} {140}},\
  \bibinfo {pages} {A1133} (\bibinfo {year} {1965})}\BibitemShut {NoStop}%
\bibitem [{\citenamefont {Cohen}\ \emph {et~al.}(2012)\citenamefont {Cohen},
  \citenamefont {Mori-S√°nchez},\ and\ \citenamefont {Yang}}]{Cohen2012}%
  \BibitemOpen
  \bibfield  {author} {\bibinfo {author} {\bibfnamefont {A.~J.}\ \bibnamefont
  {Cohen}}, \bibinfo {author} {\bibfnamefont {P.}~\bibnamefont
  {Mori-S√°nchez}}, \ and\ \bibinfo {author} {\bibfnamefont {W.}~\bibnamefont
  {Yang}},\ }\href {\doibase 10.1021/cr200107z} {\bibfield  {journal} {\bibinfo
   {journal} {Chemical Reviews}\ }\textbf {\bibinfo {volume} {112}},\ \bibinfo
  {pages} {289} (\bibinfo {year} {2012})}\BibitemShut {NoStop}%
\bibitem [{\citenamefont {Miehlich}\ \emph {et~al.}(1997)\citenamefont
  {Miehlich}, \citenamefont {Stoll},\ and\ \citenamefont
  {Savin}}]{Miechlich1997}%
  \BibitemOpen
  \bibfield  {author} {\bibinfo {author} {\bibfnamefont {B.}~\bibnamefont
  {Miehlich}}, \bibinfo {author} {\bibfnamefont {H.}~\bibnamefont {Stoll}}, \
  and\ \bibinfo {author} {\bibfnamefont {A.}~\bibnamefont {Savin}},\ }\href
  {\doibase 10.1080/002689797171418} {\bibfield  {journal} {\bibinfo  {journal}
  {Molecular Physics}\ }\textbf {\bibinfo {volume} {91}},\ \bibinfo {pages}
  {527} (\bibinfo {year} {1997})}\BibitemShut {NoStop}%
\bibitem [{\citenamefont {Grafenstein}\ \emph {et~al.}(1998)\citenamefont
  {Grafenstein}, \citenamefont {Kraka},\ and\ \citenamefont
  {Cremer}}]{Grafenstein1998}%
  \BibitemOpen
  \bibfield  {author} {\bibinfo {author} {\bibfnamefont {J.}~\bibnamefont
  {Grafenstein}}, \bibinfo {author} {\bibfnamefont {E.}~\bibnamefont {Kraka}},
  \ and\ \bibinfo {author} {\bibfnamefont {D.}~\bibnamefont {Cremer}},\ }\href
  {\doibase https://doi.org/10.1016/S0009-2614(98)00335-2} {\bibfield
  {journal} {\bibinfo  {journal} {Chemical Physics Letters}\ }\textbf {\bibinfo
  {volume} {288}},\ \bibinfo {pages} {593 } (\bibinfo {year}
  {1998})}\BibitemShut {NoStop}%
\bibitem [{\citenamefont {Grimme}\ and\ \citenamefont
  {Waletzke}(1999)}]{Grimme1999}%
  \BibitemOpen
  \bibfield  {author} {\bibinfo {author} {\bibfnamefont {S.}~\bibnamefont
  {Grimme}}\ and\ \bibinfo {author} {\bibfnamefont {M.}~\bibnamefont
  {Waletzke}},\ }\href {\doibase 10.1063/1.479866} {\bibfield  {journal}
  {\bibinfo  {journal} {The Journal of Chemical Physics}\ }\textbf {\bibinfo
  {volume} {111}},\ \bibinfo {pages} {5645} (\bibinfo {year}
  {1999})}\BibitemShut {NoStop}%
\bibitem [{\citenamefont {Wu}\ and\ \citenamefont {Shaik}(1999)}]{Wu1999}%
  \BibitemOpen
  \bibfield  {author} {\bibinfo {author} {\bibfnamefont {W.}~\bibnamefont
  {Wu}}\ and\ \bibinfo {author} {\bibfnamefont {S.}~\bibnamefont {Shaik}},\
  }\href {\doibase https://doi.org/10.1016/S0009-2614(99)00011-1} {\bibfield
  {journal} {\bibinfo  {journal} {Chemical Physics Letters}\ }\textbf {\bibinfo
  {volume} {301}},\ \bibinfo {pages} {37 } (\bibinfo {year}
  {1999})}\BibitemShut {NoStop}%
\bibitem [{\citenamefont {Zhao}\ \emph {et~al.}(2005)\citenamefont {Zhao},
  \citenamefont {Lynch},\ and\ \citenamefont {Truhlar}}]{Zhao2005}%
  \BibitemOpen
  \bibfield  {author} {\bibinfo {author} {\bibfnamefont {Y.}~\bibnamefont
  {Zhao}}, \bibinfo {author} {\bibfnamefont {B.~J.}\ \bibnamefont {Lynch}}, \
  and\ \bibinfo {author} {\bibfnamefont {D.~G.}\ \bibnamefont {Truhlar}},\
  }\href {\doibase 10.1039/B416937A} {\bibfield  {journal} {\bibinfo  {journal}
  {Phys. Chem. Chem. Phys.}\ }\textbf {\bibinfo {volume} {7}},\ \bibinfo
  {pages} {43} (\bibinfo {year} {2005})}\BibitemShut {NoStop}%
\bibitem [{\citenamefont {Ukai}\ \emph {et~al.}(2007)\citenamefont {Ukai},
  \citenamefont {Nakata}, \citenamefont {Yamanaka}, \citenamefont {Takada},\
  and\ \citenamefont {Yamaguchi}}]{Ukai2007}%
  \BibitemOpen
  \bibfield  {author} {\bibinfo {author} {\bibfnamefont {T.}~\bibnamefont
  {Ukai}}, \bibinfo {author} {\bibfnamefont {K.}~\bibnamefont {Nakata}},
  \bibinfo {author} {\bibfnamefont {S.}~\bibnamefont {Yamanaka}}, \bibinfo
  {author} {\bibfnamefont {T.}~\bibnamefont {Takada}}, \ and\ \bibinfo {author}
  {\bibfnamefont {K.}~\bibnamefont {Yamaguchi}},\ }\href {\doibase
  10.1080/00268970701618440} {\bibfield  {journal} {\bibinfo  {journal}
  {Molecular Physics}\ }\textbf {\bibinfo {volume} {105}},\ \bibinfo {pages}
  {2667} (\bibinfo {year} {2007})}\BibitemShut {NoStop}%
\bibitem [{\citenamefont {Fromager}\ \emph {et~al.}(2007)\citenamefont
  {Fromager}, \citenamefont {Toulouse},\ and\ \citenamefont
  {Jensen}}]{Fromager_2007}%
  \BibitemOpen
  \bibfield  {author} {\bibinfo {author} {\bibfnamefont {E.}~\bibnamefont
  {Fromager}}, \bibinfo {author} {\bibfnamefont {J.}~\bibnamefont {Toulouse}},
  \ and\ \bibinfo {author} {\bibfnamefont {H.~J.~A.}\ \bibnamefont {Jensen}},\
  }\href {\doibase 10.1063/1.2566459} {\bibfield  {journal} {\bibinfo
  {journal} {The Journal of Chemical Physics}\ }\textbf {\bibinfo {volume}
  {126}},\ \bibinfo {pages} {074111} (\bibinfo {year} {2007})}\BibitemShut
  {NoStop}%
\bibitem [{\citenamefont {Ying}\ \emph {et~al.}(2012)\citenamefont {Ying},
  \citenamefont {Su}, \citenamefont {Chen}, \citenamefont {Shaik},\ and\
  \citenamefont {Wu}}]{Ying2012}%
  \BibitemOpen
  \bibfield  {author} {\bibinfo {author} {\bibfnamefont {F.}~\bibnamefont
  {Ying}}, \bibinfo {author} {\bibfnamefont {P.}~\bibnamefont {Su}}, \bibinfo
  {author} {\bibfnamefont {Z.}~\bibnamefont {Chen}}, \bibinfo {author}
  {\bibfnamefont {S.}~\bibnamefont {Shaik}}, \ and\ \bibinfo {author}
  {\bibfnamefont {W.}~\bibnamefont {Wu}},\ }\href {\doibase 10.1021/ct200803h}
  {\bibfield  {journal} {\bibinfo  {journal} {Journal of Chemical Theory and
  Computation}\ }\textbf {\bibinfo {volume} {8}},\ \bibinfo {pages} {1608}
  (\bibinfo {year} {2012})}\BibitemShut {NoStop}%
\bibitem [{\citenamefont {Roemelt}\ \emph {et~al.}(2013)\citenamefont
  {Roemelt}, \citenamefont {Maganas}, \citenamefont {DeBeer},\ and\
  \citenamefont {Neese}}]{Roemelt2013}%
  \BibitemOpen
  \bibfield  {author} {\bibinfo {author} {\bibfnamefont {M.}~\bibnamefont
  {Roemelt}}, \bibinfo {author} {\bibfnamefont {D.}~\bibnamefont {Maganas}},
  \bibinfo {author} {\bibfnamefont {S.}~\bibnamefont {DeBeer}}, \ and\ \bibinfo
  {author} {\bibfnamefont {F.}~\bibnamefont {Neese}},\ }\href {\doibase
  10.1063/1.4804607} {\bibfield  {journal} {\bibinfo  {journal} {The Journal of
  Chemical Physics}\ }\textbf {\bibinfo {volume} {138}},\ \bibinfo {pages}
  {204101} (\bibinfo {year} {2013})}\BibitemShut {NoStop}%
\bibitem [{\citenamefont {Li~Manni}\ \emph {et~al.}(2014)\citenamefont
  {Li~Manni}, \citenamefont {Carlson}, \citenamefont {Luo}, \citenamefont {Ma},
  \citenamefont {Olsen}, \citenamefont {Truhlar},\ and\ \citenamefont
  {Gagliardi}}]{LiManni2014}%
  \BibitemOpen
  \bibfield  {author} {\bibinfo {author} {\bibfnamefont {G.}~\bibnamefont
  {Li~Manni}}, \bibinfo {author} {\bibfnamefont {R.~K.}\ \bibnamefont
  {Carlson}}, \bibinfo {author} {\bibfnamefont {S.}~\bibnamefont {Luo}},
  \bibinfo {author} {\bibfnamefont {D.}~\bibnamefont {Ma}}, \bibinfo {author}
  {\bibfnamefont {J.}~\bibnamefont {Olsen}}, \bibinfo {author} {\bibfnamefont
  {D.~G.}\ \bibnamefont {Truhlar}}, \ and\ \bibinfo {author} {\bibfnamefont
  {L.}~\bibnamefont {Gagliardi}},\ }\href {\doibase 10.1021/ct500483t}
  {\bibfield  {journal} {\bibinfo  {journal} {Journal of Chemical Theory and
  Computation}\ }\textbf {\bibinfo {volume} {10}},\ \bibinfo {pages} {3669}
  (\bibinfo {year} {2014})}\BibitemShut {NoStop}%
\bibitem [{\citenamefont {Roos}(1987)}]{Roos1987}%
  \BibitemOpen
  \bibfield  {author} {\bibinfo {author} {\bibfnamefont {B.~O.}\ \bibnamefont
  {Roos}},\ }\enquote {\bibinfo {title} {The complete active space
  self-consistent field method and its applications in electronic structure
  calculations},}\ in\ \href {\doibase 10.1002/9780470142943.ch1} {\emph
  {\bibinfo {booktitle} {Advances in Chemical Physics}}}\ (\bibinfo
  {publisher} {John Wiley and Sons, Ltd},\ \bibinfo {address} {New York},\
  \bibinfo {year} {1987})\ pp.\ \bibinfo {pages} {399--445}\BibitemShut
  {NoStop}%
\bibitem [{\citenamefont {Shepard}(1987)}]{Shepard1987}%
  \BibitemOpen
  \bibfield  {author} {\bibinfo {author} {\bibfnamefont {R.}~\bibnamefont
  {Shepard}},\ }\enquote {\bibinfo {title} {The multiconfiguration
  self-consistent field method},}\ in\ \href {\doibase
  10.1002/9780470142943.ch1} {\emph {\bibinfo {booktitle} {Advances in Chemical
  Physics}}}\ (\bibinfo  {publisher} {John Wiley and Sons, Ltd},\ \bibinfo
  {address} {New York},\ \bibinfo {year} {1987})\ pp.\ \bibinfo {pages}
  {63--200}\BibitemShut {NoStop}%
\bibitem [{\citenamefont {Sherrill}\ and\ \citenamefont
  {Schaefer}(1999)}]{Lowdin1999}%
  \BibitemOpen
  \bibfield  {author} {\bibinfo {author} {\bibfnamefont {C.~D.}\ \bibnamefont
  {Sherrill}}\ and\ \bibinfo {author} {\bibfnamefont {H.~F.}\ \bibnamefont
  {Schaefer}}\ }(\bibinfo  {publisher} {Academic Press},\ \bibinfo {year}
  {1999})\ pp.\ \bibinfo {pages} {143 -- 269}\BibitemShut {NoStop}%
\bibitem [{\citenamefont {Werner}(1987)}]{Werner2007}%
  \BibitemOpen
  \bibfield  {author} {\bibinfo {author} {\bibfnamefont {H.-J.}\ \bibnamefont
  {Werner}},\ }\enquote {\bibinfo {title} {Matrix-formulated direct
  multiconfiguration self-consistent field and multiconfiguration reference
  configuration-interaction methods},}\ in\ \href {\doibase
  10.1002/9780470142943.ch1} {\emph {\bibinfo {booktitle} {Advances in Chemical
  Physics}}}\ (\bibinfo  {publisher} {John Wiley and Sons, Ltd},\ \bibinfo
  {year} {1987})\ pp.\ \bibinfo {pages} {1--62}\BibitemShut {NoStop}%
\bibitem [{\citenamefont {Shamasundar}\ \emph {et~al.}(2011)\citenamefont
  {Shamasundar}, \citenamefont {Knizia},\ and\ \citenamefont
  {Werner}}]{Shamasundar2011}%
  \BibitemOpen
  \bibfield  {author} {\bibinfo {author} {\bibfnamefont {K.~R.}\ \bibnamefont
  {Shamasundar}}, \bibinfo {author} {\bibfnamefont {G.}~\bibnamefont {Knizia}},
  \ and\ \bibinfo {author} {\bibfnamefont {H.-J.}\ \bibnamefont {Werner}},\
  }\href {\doibase 10.1063/1.3609809} {\bibfield  {journal} {\bibinfo
  {journal} {The Journal of Chemical Physics}\ }\textbf {\bibinfo {volume}
  {135}},\ \bibinfo {pages} {054101} (\bibinfo {year} {2011})}\BibitemShut
  {NoStop}%
\bibitem [{\citenamefont {Szalay}\ \emph {et~al.}(2012)\citenamefont {Szalay},
  \citenamefont {Muller}, \citenamefont {Gidofalvi}, \citenamefont {Lischka},\
  and\ \citenamefont {Shepard}}]{Szalay2012}%
  \BibitemOpen
  \bibfield  {author} {\bibinfo {author} {\bibfnamefont {P.~G.}\ \bibnamefont
  {Szalay}}, \bibinfo {author} {\bibfnamefont {T.}~\bibnamefont {Muller}},
  \bibinfo {author} {\bibfnamefont {G.}~\bibnamefont {Gidofalvi}}, \bibinfo
  {author} {\bibfnamefont {H.}~\bibnamefont {Lischka}}, \ and\ \bibinfo
  {author} {\bibfnamefont {R.}~\bibnamefont {Shepard}},\ }\href {\doibase
  10.1021/cr200137a} {\bibfield  {journal} {\bibinfo  {journal} {Chemical
  Reviews}\ }\textbf {\bibinfo {volume} {112}},\ \bibinfo {pages} {108}
  (\bibinfo {year} {2012})}\BibitemShut {NoStop}%
\bibitem [{\citenamefont {Helgaker}\ \emph {et~al.}(2014)\citenamefont
  {Helgaker}, \citenamefont {Jorgensen},\ and\ \citenamefont
  {Olsen}}]{Helgaker2014}%
  \BibitemOpen
  \bibfield  {author} {\bibinfo {author} {\bibfnamefont {T.}~\bibnamefont
  {Helgaker}}, \bibinfo {author} {\bibfnamefont {P.}~\bibnamefont {Jorgensen}},
  \ and\ \bibinfo {author} {\bibfnamefont {J.}~\bibnamefont {Olsen}},\
  }\enquote {\bibinfo {title} {Configuration-interaction theory},}\ in\ \href
  {\doibase 10.1002/9781119019572.ch11} {\emph {\bibinfo {booktitle} {Molecular
  Electronic Structure Theory}}}\ (\bibinfo  {publisher} {John Wiley and Sons,
  Ltd},\ \bibinfo {year} {2014})\ Chap.~\bibinfo {chapter} {11}, pp.\ \bibinfo
  {pages} {523--597}\BibitemShut {NoStop}%
\bibitem [{\citenamefont {Lischka}\ \emph {et~al.}(2018)\citenamefont
  {Lischka}, \citenamefont {Nachtigallova}, \citenamefont {Aquino},
  \citenamefont {Szalay}, \citenamefont {Plasser}, \citenamefont {Machado},\
  and\ \citenamefont {Barbatti}}]{Lischka2018}%
  \BibitemOpen
  \bibfield  {author} {\bibinfo {author} {\bibfnamefont {H.}~\bibnamefont
  {Lischka}}, \bibinfo {author} {\bibfnamefont {D.}~\bibnamefont
  {Nachtigallova}}, \bibinfo {author} {\bibfnamefont {A.~J.~A.}\ \bibnamefont
  {Aquino}}, \bibinfo {author} {\bibfnamefont {P.~G.}\ \bibnamefont {Szalay}},
  \bibinfo {author} {\bibfnamefont {F.}~\bibnamefont {Plasser}}, \bibinfo
  {author} {\bibfnamefont {F.~B.~C.}\ \bibnamefont {Machado}}, \ and\ \bibinfo
  {author} {\bibfnamefont {M.}~\bibnamefont {Barbatti}},\ }\href {\doibase
  10.1021/acs.chemrev.8b00244} {\bibfield  {journal} {\bibinfo  {journal}
  {Chemical Reviews}\ }\textbf {\bibinfo {volume} {118}},\ \bibinfo {pages}
  {7293} (\bibinfo {year} {2018})}\BibitemShut {NoStop}%
\bibitem [{\citenamefont {White}(1992)}]{White-1992b}%
  \BibitemOpen
  \bibfield  {author} {\bibinfo {author} {\bibfnamefont {S.~R.}\ \bibnamefont
  {White}},\ }\href {\doibase 10.1103/PhysRevLett.69.2863} {\bibfield
  {journal} {\bibinfo  {journal} {Phys. Rev. Lett.}\ }\textbf {\bibinfo
  {volume} {69}},\ \bibinfo {pages} {2863} (\bibinfo {year}
  {1992})}\BibitemShut {NoStop}%
\bibitem [{\citenamefont {White}(1993)}]{White-1993}%
  \BibitemOpen
  \bibfield  {author} {\bibinfo {author} {\bibfnamefont {S.~R.}\ \bibnamefont
  {White}},\ }\href {\doibase 10.1103/PhysRevB.48.10345} {\bibfield  {journal}
  {\bibinfo  {journal} {Phys. Rev. B}\ }\textbf {\bibinfo {volume} {48}},\
  \bibinfo {pages} {10345} (\bibinfo {year} {1993})}\BibitemShut {NoStop}%
\bibitem [{\citenamefont {Xiang}(1996)}]{Xiang-1996}%
  \BibitemOpen
  \bibfield  {author} {\bibinfo {author} {\bibfnamefont {T.}~\bibnamefont
  {Xiang}},\ }\href {\doibase 10.1103/PhysRevB.53.R10445} {\bibfield  {journal}
  {\bibinfo  {journal} {Phys. Rev. B}\ }\textbf {\bibinfo {volume} {53}},\
  \bibinfo {pages} {R10445} (\bibinfo {year} {1996})}\BibitemShut {NoStop}%
\bibitem [{\citenamefont {Nishimoto}\ \emph {et~al.}(2002)\citenamefont
  {Nishimoto}, \citenamefont {Jeckelmann}, \citenamefont {Gebhard},\ and\
  \citenamefont {Noack}}]{Nishimoto-2002}%
  \BibitemOpen
  \bibfield  {author} {\bibinfo {author} {\bibfnamefont {S.}~\bibnamefont
  {Nishimoto}}, \bibinfo {author} {\bibfnamefont {E.}~\bibnamefont
  {Jeckelmann}}, \bibinfo {author} {\bibfnamefont {F.}~\bibnamefont {Gebhard}},
  \ and\ \bibinfo {author} {\bibfnamefont {R.~M.}\ \bibnamefont {Noack}},\
  }\href {\doibase 10.1103/PhysRevB.65.165114} {\bibfield  {journal} {\bibinfo
  {journal} {Phys. Rev. B}\ }\textbf {\bibinfo {volume} {65}},\ \bibinfo
  {pages} {165114} (\bibinfo {year} {2002})}\BibitemShut {NoStop}%
\bibitem [{\citenamefont {Legeza}\ and\ \citenamefont
  {S{\'o}lyom}(2003)}]{Legeza-2003b}%
  \BibitemOpen
  \bibfield  {author} {\bibinfo {author} {\bibfnamefont {{\"O}.}~\bibnamefont
  {Legeza}}\ and\ \bibinfo {author} {\bibfnamefont {J.}~\bibnamefont
  {S{\'o}lyom}},\ }\href {\doibase 10.1103/PhysRevB.68.195116} {\bibfield
  {journal} {\bibinfo  {journal} {Phys. Rev. B}\ }\textbf {\bibinfo {volume}
  {68}},\ \bibinfo {pages} {195116} (\bibinfo {year} {2003})}\BibitemShut
  {NoStop}%
\bibitem [{\citenamefont {White}\ and\ \citenamefont
  {Martin}(1999)}]{White-1999}%
  \BibitemOpen
  \bibfield  {author} {\bibinfo {author} {\bibfnamefont {S.~R.}\ \bibnamefont
  {White}}\ and\ \bibinfo {author} {\bibfnamefont {R.~L.}\ \bibnamefont
  {Martin}},\ }\href {\doibase http://dx.doi.org/10.1063/1.478295} {\bibfield
  {journal} {\bibinfo  {journal} {J. Chem. Phys.}\ }\textbf {\bibinfo {volume}
  {110}},\ \bibinfo {pages} {4127} (\bibinfo {year} {1999})}\BibitemShut
  {NoStop}%
\bibitem [{\citenamefont {Chan}\ and\ \citenamefont
  {Head-Gordon}(2002)}]{Chan-2002a}%
  \BibitemOpen
  \bibfield  {author} {\bibinfo {author} {\bibfnamefont {G.~K.-L.}\
  \bibnamefont {Chan}}\ and\ \bibinfo {author} {\bibfnamefont {M.}~\bibnamefont
  {Head-Gordon}},\ }\href {\doibase http://dx.doi.org/10.1063/1.1449459}
  {\bibfield  {journal} {\bibinfo  {journal} {J. Chem. Phys.}\ }\textbf
  {\bibinfo {volume} {116}},\ \bibinfo {pages} {4462} (\bibinfo {year}
  {2002})}\BibitemShut {NoStop}%
\bibitem [{\citenamefont {Legeza}\ \emph {et~al.}(2003)\citenamefont {Legeza},
  \citenamefont {R\"oder},\ and\ \citenamefont {Hess}}]{Legeza-2003a}%
  \BibitemOpen
  \bibfield  {author} {\bibinfo {author} {\bibfnamefont {{\"O}.}~\bibnamefont
  {Legeza}}, \bibinfo {author} {\bibfnamefont {J.}~\bibnamefont {R\"oder}}, \
  and\ \bibinfo {author} {\bibfnamefont {B.~A.}\ \bibnamefont {Hess}},\ }\href
  {\doibase 10.1103/PhysRevB.67.125114} {\bibfield  {journal} {\bibinfo
  {journal} {Phys. Rev. B}\ }\textbf {\bibinfo {volume} {67}},\ \bibinfo
  {pages} {125114} (\bibinfo {year} {2003})}\BibitemShut {NoStop}%
\bibitem [{\citenamefont {Rissler}\ \emph {et~al.}(2006)\citenamefont
  {Rissler}, \citenamefont {Noack},\ and\ \citenamefont {White}}]{Rissler2006}%
  \BibitemOpen
  \bibfield  {author} {\bibinfo {author} {\bibfnamefont {J.}~\bibnamefont
  {Rissler}}, \bibinfo {author} {\bibfnamefont {R.~M.}\ \bibnamefont {Noack}},
  \ and\ \bibinfo {author} {\bibfnamefont {S.~R.}\ \bibnamefont {White}},\
  }\href {\doibase https://doi.org/10.1016/j.chemphys.2005.10.018} {\bibfield
  {journal} {\bibinfo  {journal} {Chemical Physics}\ }\textbf {\bibinfo
  {volume} {323}},\ \bibinfo {pages} {519 } (\bibinfo {year}
  {2006})}\BibitemShut {NoStop}%
\bibitem [{\citenamefont {Barcza}\ \emph {et~al.}(2011)\citenamefont {Barcza},
  \citenamefont {Legeza}, \citenamefont {Marti},\ and\ \citenamefont
  {Reiher}}]{Barcza-2011}%
  \BibitemOpen
  \bibfield  {author} {\bibinfo {author} {\bibfnamefont {G.}~\bibnamefont
  {Barcza}}, \bibinfo {author} {\bibfnamefont {{\"O}.}~\bibnamefont {Legeza}},
  \bibinfo {author} {\bibfnamefont {K.~H.}\ \bibnamefont {Marti}}, \ and\
  \bibinfo {author} {\bibfnamefont {M.}~\bibnamefont {Reiher}},\ }\href
  {\doibase 10.1103/PhysRevA.83.012508} {\bibfield  {journal} {\bibinfo
  {journal} {Phys. Rev. A}\ }\textbf {\bibinfo {volume} {83}},\ \bibinfo
  {pages} {012508} (\bibinfo {year} {2011})}\BibitemShut {NoStop}%
\bibitem [{\citenamefont {{\relax Sz}alay}\ \emph {et~al.}(2015)\citenamefont
  {{\relax Sz}alay}, \citenamefont {Pfeffer}, \citenamefont {Murg},
  \citenamefont {Barcza}, \citenamefont {Verstraete}, \citenamefont
  {Schneider},\ and\ \citenamefont {Legeza}}]{Szalay-2015a}%
  \BibitemOpen
  \bibfield  {author} {\bibinfo {author} {\bibfnamefont {{\relax
  Sz}.}~\bibnamefont {{\relax Sz}alay}}, \bibinfo {author} {\bibfnamefont
  {M.}~\bibnamefont {Pfeffer}}, \bibinfo {author} {\bibfnamefont
  {V.}~\bibnamefont {Murg}}, \bibinfo {author} {\bibfnamefont {G.}~\bibnamefont
  {Barcza}}, \bibinfo {author} {\bibfnamefont {F.}~\bibnamefont {Verstraete}},
  \bibinfo {author} {\bibfnamefont {R.}~\bibnamefont {Schneider}}, \ and\
  \bibinfo {author} {\bibfnamefont {{\"O}.}~\bibnamefont {Legeza}},\ }\href
  {\doibase 10.1002/qua.24898} {\bibfield  {journal} {\bibinfo  {journal} {Int.
  J. Quant. Chem.}\ }\textbf {\bibinfo {volume} {115}},\ \bibinfo {pages}
  {1342} (\bibinfo {year} {2015})}\BibitemShut {NoStop}%
\bibitem [{\citenamefont {Schollw\"ock}(2005)}]{Schollwock2005}%
  \BibitemOpen
  \bibfield  {author} {\bibinfo {author} {\bibfnamefont {U.}~\bibnamefont
  {Schollw\"ock}},\ }\href {\doibase 10.1103/RevModPhys.77.259} {\bibfield
  {journal} {\bibinfo  {journal} {Rev. Mod. Phys.}\ }\textbf {\bibinfo {volume}
  {77}},\ \bibinfo {pages} {259} (\bibinfo {year} {2005})}\BibitemShut
  {NoStop}%
\bibitem [{\citenamefont {Schollw\"ock}(2011)}]{Schollwock-2011}%
  \BibitemOpen
  \bibfield  {author} {\bibinfo {author} {\bibfnamefont {U.}~\bibnamefont
  {Schollw\"ock}},\ }\href {\doibase https://doi.org/10.1016/j.aop.2010.09.012}
  {\bibfield  {journal} {\bibinfo  {journal} {Annals of Physics}\ }\textbf
  {\bibinfo {volume} {326}},\ \bibinfo {pages} {96 } (\bibinfo {year}
  {2011})},\ \bibinfo {note} {january 2011 Special Issue}\BibitemShut {NoStop}%
\bibitem [{\citenamefont {Olivares-Amaya}\ \emph {et~al.}(2015)\citenamefont
  {Olivares-Amaya}, \citenamefont {Hu}, \citenamefont {Nakatani}, \citenamefont
  {Sharma}, \citenamefont {Yang},\ and\ \citenamefont {Chan}}]{Olivares-2015}%
  \BibitemOpen
  \bibfield  {author} {\bibinfo {author} {\bibfnamefont {R.}~\bibnamefont
  {Olivares-Amaya}}, \bibinfo {author} {\bibfnamefont {W.}~\bibnamefont {Hu}},
  \bibinfo {author} {\bibfnamefont {N.}~\bibnamefont {Nakatani}}, \bibinfo
  {author} {\bibfnamefont {S.}~\bibnamefont {Sharma}}, \bibinfo {author}
  {\bibfnamefont {J.}~\bibnamefont {Yang}}, \ and\ \bibinfo {author}
  {\bibfnamefont {G.~K.-L.}\ \bibnamefont {Chan}},\ }\href
  {https://aip.scitation.org/doi/10.1063/1.4905329} {\bibfield  {journal}
  {\bibinfo  {journal} {The Journal of Chemical Physics}\ }\textbf {\bibinfo
  {volume} {142}} (\bibinfo {year} {2015})}\BibitemShut {NoStop}%
\bibitem [{\citenamefont {Baiardi}\ and\ \citenamefont
  {Reiher}(2020)}]{Baiardi2020}%
  \BibitemOpen
  \bibfield  {author} {\bibinfo {author} {\bibfnamefont {A.}~\bibnamefont
  {Baiardi}}\ and\ \bibinfo {author} {\bibfnamefont {M.}~\bibnamefont
  {Reiher}},\ }\href {\doibase 10.1063/1.5129672} {\bibfield  {journal}
  {\bibinfo  {journal} {The Journal of Chemical Physics}\ }\textbf {\bibinfo
  {volume} {152}},\ \bibinfo {pages} {040903} (\bibinfo {year} {2020})},\
  \Eprint {http://arxiv.org/abs/https://doi.org/10.1063/1.5129672}
  {https://doi.org/10.1063/1.5129672} \BibitemShut {NoStop}%
\bibitem [{\citenamefont {Dukelsky}\ \emph {et~al.}(2002)\citenamefont
  {Dukelsky}, \citenamefont {Pittel}, \citenamefont {Dimitrova},\ and\
  \citenamefont {Stoitsov}}]{Dukelsky-2002}%
  \BibitemOpen
  \bibfield  {author} {\bibinfo {author} {\bibfnamefont {J.}~\bibnamefont
  {Dukelsky}}, \bibinfo {author} {\bibfnamefont {S.}~\bibnamefont {Pittel}},
  \bibinfo {author} {\bibfnamefont {S.~S.}\ \bibnamefont {Dimitrova}}, \ and\
  \bibinfo {author} {\bibfnamefont {M.~V.}\ \bibnamefont {Stoitsov}},\ }\href
  {\doibase 10.1103/PhysRevC.65.054319} {\bibfield  {journal} {\bibinfo
  {journal} {Phys. Rev. C}\ }\textbf {\bibinfo {volume} {65}},\ \bibinfo
  {pages} {054319} (\bibinfo {year} {2002})}\BibitemShut {NoStop}%
\bibitem [{\citenamefont {Papenbrock}\ and\ \citenamefont
  {Dean}(2005)}]{Papenbrock2005}%
  \BibitemOpen
  \bibfield  {author} {\bibinfo {author} {\bibfnamefont {T.}~\bibnamefont
  {Papenbrock}}\ and\ \bibinfo {author} {\bibfnamefont {D.~J.}\ \bibnamefont
  {Dean}},\ }\href {\doibase 10.1088/0954-3899/31/8/016} {\bibfield  {journal}
  {\bibinfo  {journal} {Journal of Physics G: Nuclear and Particle Physics}\
  }\textbf {\bibinfo {volume} {31}},\ \bibinfo {pages} {S1377} (\bibinfo {year}
  {2005})}\BibitemShut {NoStop}%
\bibitem [{\citenamefont {Rotureau}\ \emph {et~al.}(2006)\citenamefont
  {Rotureau}, \citenamefont {Michel}, \citenamefont {Nazarewicz}, \citenamefont
  {P\l{}oszajczak},\ and\ \citenamefont {Dukelsky}}]{Rotureau2006}%
  \BibitemOpen
  \bibfield  {author} {\bibinfo {author} {\bibfnamefont {J.}~\bibnamefont
  {Rotureau}}, \bibinfo {author} {\bibfnamefont {N.}~\bibnamefont {Michel}},
  \bibinfo {author} {\bibfnamefont {W.}~\bibnamefont {Nazarewicz}}, \bibinfo
  {author} {\bibfnamefont {M.}~\bibnamefont {P\l{}oszajczak}}, \ and\ \bibinfo
  {author} {\bibfnamefont {J.}~\bibnamefont {Dukelsky}},\ }\href
  {https://link.aps.org/doi/10.1103/PhysRevLett.97.110603} {\bibfield
  {journal} {\bibinfo  {journal} {Phys. Rev. Lett.}\ }\textbf {\bibinfo
  {volume} {97}},\ \bibinfo {pages} {110603} (\bibinfo {year}
  {2006})}\BibitemShut {NoStop}%
\bibitem [{\citenamefont {Legeza}\ \emph {et~al.}(2015)\citenamefont {Legeza},
  \citenamefont {Veis}, \citenamefont {Poves},\ and\ \citenamefont
  {Dukelsky}}]{Legeza-2015}%
  \BibitemOpen
  \bibfield  {author} {\bibinfo {author} {\bibfnamefont {{\"O}.}~\bibnamefont
  {Legeza}}, \bibinfo {author} {\bibfnamefont {L.}~\bibnamefont {Veis}},
  \bibinfo {author} {\bibfnamefont {A.}~\bibnamefont {Poves}}, \ and\ \bibinfo
  {author} {\bibfnamefont {J.}~\bibnamefont {Dukelsky}},\ }\href {\doibase
  10.1103/PhysRevC.92.051303} {\bibfield  {journal} {\bibinfo  {journal} {Phys.
  Rev. C}\ }\textbf {\bibinfo {volume} {92}},\ \bibinfo {pages} {051303}
  (\bibinfo {year} {2015})}\BibitemShut {NoStop}%
\bibitem [{\citenamefont {Legeza}\ and\ \citenamefont
  {Schilling}(2018)}]{Legeza-2018a}%
  \BibitemOpen
  \bibfield  {author} {\bibinfo {author} {\bibfnamefont {O.}~\bibnamefont
  {Legeza}}\ and\ \bibinfo {author} {\bibfnamefont {C.}~\bibnamefont
  {Schilling}},\ }\href {\doibase 10.1103/PhysRevA.97.052105} {\bibfield
  {journal} {\bibinfo  {journal} {Phys. Rev. A}\ }\textbf {\bibinfo {volume}
  {97}},\ \bibinfo {pages} {052105} (\bibinfo {year} {2018})}\BibitemShut
  {NoStop}%
\bibitem [{\citenamefont {Shapir}\ \emph {et~al.}(2019)\citenamefont {Shapir},
  \citenamefont {Hamo}, \citenamefont {Pecker}, \citenamefont {Moca},
  \citenamefont {Legeza}, \citenamefont {Zarand},\ and\ \citenamefont
  {Ilani}}]{Shapir-2019}%
  \BibitemOpen
  \bibfield  {author} {\bibinfo {author} {\bibfnamefont {I.}~\bibnamefont
  {Shapir}}, \bibinfo {author} {\bibfnamefont {A.}~\bibnamefont {Hamo}},
  \bibinfo {author} {\bibfnamefont {S.}~\bibnamefont {Pecker}}, \bibinfo
  {author} {\bibfnamefont {C.~P.}\ \bibnamefont {Moca}}, \bibinfo {author}
  {\bibfnamefont {O.}~\bibnamefont {Legeza}}, \bibinfo {author} {\bibfnamefont
  {G.}~\bibnamefont {Zarand}}, \ and\ \bibinfo {author} {\bibfnamefont
  {S.}~\bibnamefont {Ilani}},\ }\href {\doibase 10.1126/science.aat0905}
  {\bibfield  {journal} {\bibinfo  {journal} {Science}\ }\textbf {\bibinfo
  {volume} {364}},\ \bibinfo {pages} {870} (\bibinfo {year}
  {2019})}\BibitemShut {NoStop}%
\bibitem [{\citenamefont {Moca}\ \emph {et~al.}(2019)\citenamefont {Moca},
  \citenamefont {Izumida}, \citenamefont {Dora}, \citenamefont {Legeza},\ and\
  \citenamefont {Zarand}}]{Moca-2019}%
  \BibitemOpen
  \bibfield  {author} {\bibinfo {author} {\bibfnamefont {C.~P.}\ \bibnamefont
  {Moca}}, \bibinfo {author} {\bibfnamefont {W.}~\bibnamefont {Izumida}},
  \bibinfo {author} {\bibfnamefont {B.}~\bibnamefont {Dora}}, \bibinfo {author}
  {\bibfnamefont {O.}~\bibnamefont {Legeza}}, \ and\ \bibinfo {author}
  {\bibfnamefont {G.}~\bibnamefont {Zarand}},\ }\href@noop {} {\enquote
  {\bibinfo {title} {Topologically protected, correlated end spin formation in
  carbon nanotubes},}\ } (\bibinfo {year} {2019}),\ \Eprint
  {http://arxiv.org/abs/1909.08400} {arXiv:1909.08400} \BibitemShut {NoStop}%
\bibitem [{\citenamefont {Knecht}\ \emph {et~al.}(2014)\citenamefont {Knecht},
  \citenamefont {Legeza},\ and\ \citenamefont {Reiher}}]{Knecht2014}%
  \BibitemOpen
  \bibfield  {author} {\bibinfo {author} {\bibfnamefont {S.}~\bibnamefont
  {Knecht}}, \bibinfo {author} {\bibfnamefont {√.}~\bibnamefont {Legeza}}, \
  and\ \bibinfo {author} {\bibfnamefont {M.}~\bibnamefont {Reiher}},\ }\href
  {https://doi.org/10.1063/1.4862495} {\bibfield  {journal} {\bibinfo
  {journal} {The Journal of Chemical Physics}\ }\textbf {\bibinfo {volume}
  {140}},\ \bibinfo {pages} {041101} (\bibinfo {year} {2014})}\BibitemShut
  {NoStop}%
\bibitem [{\citenamefont {Battaglia}\ \emph {et~al.}(2018)\citenamefont
  {Battaglia}, \citenamefont {Keller},\ and\ \citenamefont
  {Knecht}}]{Battaglia2018}%
  \BibitemOpen
  \bibfield  {author} {\bibinfo {author} {\bibfnamefont {S.}~\bibnamefont
  {Battaglia}}, \bibinfo {author} {\bibfnamefont {S.}~\bibnamefont {Keller}}, \
  and\ \bibinfo {author} {\bibfnamefont {S.}~\bibnamefont {Knecht}},\ }\href
  {https://doi.org/10.1021/acs.jctc.7b01065} {\bibfield  {journal} {\bibinfo
  {journal} {Journal of Chemical Theory and Computation}\ }\textbf {\bibinfo
  {volume} {14}},\ \bibinfo {pages} {2353} (\bibinfo {year}
  {2018})}\BibitemShut {NoStop}%
\bibitem [{\citenamefont {Brandejs}\ \emph {et~al.}(2020)\citenamefont
  {Brandejs}, \citenamefont {Visnak}, \citenamefont {Veis}, \citenamefont
  {Mate}, \citenamefont {Legeza},\ and\ \citenamefont
  {Pittner}}]{Brandejs2020}%
  \BibitemOpen
  \bibfield  {author} {\bibinfo {author} {\bibfnamefont {J.}~\bibnamefont
  {Brandejs}}, \bibinfo {author} {\bibfnamefont {J.}~\bibnamefont {Visnak}},
  \bibinfo {author} {\bibfnamefont {L.}~\bibnamefont {Veis}}, \bibinfo {author}
  {\bibfnamefont {M.}~\bibnamefont {Mate}}, \bibinfo {author} {\bibfnamefont
  {O.}~\bibnamefont {Legeza}}, \ and\ \bibinfo {author} {\bibfnamefont
  {J.}~\bibnamefont {Pittner}},\ }\href {https://doi.org/10.1063/1.5144974}
  {\bibfield  {journal} {\bibinfo  {journal} {The Journal of Chemical Physics}\
  }\textbf {\bibinfo {volume} {152}},\ \bibinfo {pages} {174107} (\bibinfo
  {year} {2020})}\BibitemShut {NoStop}%
\bibitem [{\citenamefont {Cramer}(2005)}]{Cramer2005}%
  \BibitemOpen
  \bibfield  {author} {\bibinfo {author} {\bibfnamefont {C.}~\bibnamefont
  {Cramer}},\ }\href
  {https://www.wiley.com/en-us/Essentials+of+Computational+Chemistry%3A+Theories+and+Models%2C+2nd+Edition-p-9780470091821}
  {\emph {\bibinfo {title} {Essentials of Computational Chemistry: Theories and
  Models}}}\ (\bibinfo  {publisher} {Wiley},\ \bibinfo {year}
  {2005})\BibitemShut {NoStop}%
\bibitem [{\citenamefont {Jensen}(2006)}]{Jensen2006}%
  \BibitemOpen
  \bibfield  {author} {\bibinfo {author} {\bibfnamefont {F.}~\bibnamefont
  {Jensen}},\ }\href@noop {} {\emph {\bibinfo {title} {Introduction to
  Computational Chemistry}}}\ (\bibinfo  {publisher} {John Wiley and Sons,
  Inc.},\ \bibinfo {address} {Hoboken, NJ, USA},\ \bibinfo {year}
  {2006})\BibitemShut {NoStop}%
\bibitem [{\citenamefont {Sharma}\ \emph {et~al.}(2014)\citenamefont {Sharma},
  \citenamefont {Sivalingam}, \citenamefont {Neese},\ and\ \citenamefont
  {Chan}}]{Sharma2014}%
  \BibitemOpen
  \bibfield  {author} {\bibinfo {author} {\bibfnamefont {S.}~\bibnamefont
  {Sharma}}, \bibinfo {author} {\bibfnamefont {K.}~\bibnamefont {Sivalingam}},
  \bibinfo {author} {\bibfnamefont {F.}~\bibnamefont {Neese}}, \ and\ \bibinfo
  {author} {\bibfnamefont {G.~K.-L.}\ \bibnamefont {Chan}},\ }\href {\doibase
  10.1038/nchem.2041} {\bibfield  {journal} {\bibinfo  {journal} {Nature
  Chemistry}\ }\textbf {\bibinfo {volume} {6}},\ \bibinfo {pages} {927}
  (\bibinfo {year} {2014})}\BibitemShut {NoStop}%
\bibitem [{\citenamefont {Kawakami}\ \emph {et~al.}(2017)\citenamefont
  {Kawakami}, \citenamefont {Sano}, \citenamefont {Saito}, \citenamefont
  {Sharma}, \citenamefont {Shoji}, \citenamefont {Yamada}, \citenamefont
  {Takano}, \citenamefont {Yamanaka}, \citenamefont {Okumura}, \citenamefont
  {Nakajima},\ and\ \citenamefont {Yamaguchi}}]{Takashi2017}%
  \BibitemOpen
  \bibfield  {author} {\bibinfo {author} {\bibfnamefont {T.}~\bibnamefont
  {Kawakami}}, \bibinfo {author} {\bibfnamefont {S.}~\bibnamefont {Sano}},
  \bibinfo {author} {\bibfnamefont {T.}~\bibnamefont {Saito}}, \bibinfo
  {author} {\bibfnamefont {S.}~\bibnamefont {Sharma}}, \bibinfo {author}
  {\bibfnamefont {M.}~\bibnamefont {Shoji}}, \bibinfo {author} {\bibfnamefont
  {S.}~\bibnamefont {Yamada}}, \bibinfo {author} {\bibfnamefont
  {Y.}~\bibnamefont {Takano}}, \bibinfo {author} {\bibfnamefont
  {S.}~\bibnamefont {Yamanaka}}, \bibinfo {author} {\bibfnamefont
  {M.}~\bibnamefont {Okumura}}, \bibinfo {author} {\bibfnamefont
  {T.}~\bibnamefont {Nakajima}}, \ and\ \bibinfo {author} {\bibfnamefont
  {K.}~\bibnamefont {Yamaguchi}},\ }\href {\doibase
  10.1080/00268976.2017.1301586} {\bibfield  {journal} {\bibinfo  {journal}
  {Molecular Physics}\ }\textbf {\bibinfo {volume} {115}},\ \bibinfo {pages}
  {2154} (\bibinfo {year} {2017})}\BibitemShut {NoStop}%
\bibitem [{\citenamefont {Iv{\'a}dy}\ \emph {et~al.}(2020)\citenamefont
  {Iv{\'a}dy}, \citenamefont {Barcza}, \citenamefont {Thiering}, \citenamefont
  {Li}, \citenamefont {Hamdi}, \citenamefont {Chou}, \citenamefont {Legeza},\
  and\ \citenamefont {Gali}}]{Ivady-2019}%
  \BibitemOpen
  \bibfield  {author} {\bibinfo {author} {\bibfnamefont {V.}~\bibnamefont
  {Iv{\'a}dy}}, \bibinfo {author} {\bibfnamefont {G.}~\bibnamefont {Barcza}},
  \bibinfo {author} {\bibfnamefont {G.}~\bibnamefont {Thiering}}, \bibinfo
  {author} {\bibfnamefont {S.}~\bibnamefont {Li}}, \bibinfo {author}
  {\bibfnamefont {H.}~\bibnamefont {Hamdi}}, \bibinfo {author} {\bibfnamefont
  {J.-P.}\ \bibnamefont {Chou}}, \bibinfo {author} {\bibfnamefont
  {{\"O}.}~\bibnamefont {Legeza}}, \ and\ \bibinfo {author} {\bibfnamefont
  {A.}~\bibnamefont {Gali}},\ }\href {\doibase 10.1038/s41524-020-0305-x}
  {\bibfield  {journal} {\bibinfo  {journal} {npj Computational Materials}\
  }\textbf {\bibinfo {volume} {6}},\ \bibinfo {pages} {41} (\bibinfo {year}
  {2020})}\BibitemShut {NoStop}%
\bibitem [{\citenamefont {Szabo}\ and\ \citenamefont
  {Ostlund}(1996)}]{aszabo82_qchem}%
  \BibitemOpen
  \bibfield  {author} {\bibinfo {author} {\bibfnamefont {A.}~\bibnamefont
  {Szabo}}\ and\ \bibinfo {author} {\bibfnamefont {N.~S.}\ \bibnamefont
  {Ostlund}},\ }\href@noop {} {\emph {\bibinfo {title} {Modern Quantum
  Chemistry: Introduction to Advanced Electronic Structure Theory}}},\ \bibinfo
  {edition} {1st}\ ed.\ (\bibinfo  {publisher} {Dover Publications, Inc.},\
  \bibinfo {address} {Mineola},\ \bibinfo {year} {1996})\BibitemShut {NoStop}%
\bibitem [{\citenamefont {Giannozzi}\ \emph {et~al.}(2009)\citenamefont
  {Giannozzi}, \citenamefont {Baroni}, \citenamefont {Bonini}, \citenamefont
  {Calandra}, \citenamefont {Car}, \citenamefont {Cavazzoni}, \citenamefont
  {Ceresoli}, \citenamefont {Chiarotti}, \citenamefont {Cococcioni},
  \citenamefont {Dabo}, \citenamefont {Corso}, \citenamefont {de~Gironcoli},
  \citenamefont {Fabris}, \citenamefont {Fratesi}, \citenamefont {Gebauer},
  \citenamefont {Gerstmann}, \citenamefont {Gougoussis}, \citenamefont
  {Kokalj}, \citenamefont {Lazzeri}, \citenamefont {Martin-Samos},
  \citenamefont {Marzari}, \citenamefont {Mauri}, \citenamefont {Mazzarello},
  \citenamefont {Paolini}, \citenamefont {Pasquarello}, \citenamefont
  {Paulatto}, \citenamefont {Sbraccia}, \citenamefont {Scandolo}, \citenamefont
  {Sclauzero}, \citenamefont {Seitsonen}, \citenamefont {Smogunov},
  \citenamefont {Umari},\ and\ \citenamefont {Wentzcovitch}}]{Giannozzi2009}%
  \BibitemOpen
  \bibfield  {author} {\bibinfo {author} {\bibfnamefont {P.}~\bibnamefont
  {Giannozzi}}, \bibinfo {author} {\bibfnamefont {S.}~\bibnamefont {Baroni}},
  \bibinfo {author} {\bibfnamefont {N.}~\bibnamefont {Bonini}}, \bibinfo
  {author} {\bibfnamefont {M.}~\bibnamefont {Calandra}}, \bibinfo {author}
  {\bibfnamefont {R.}~\bibnamefont {Car}}, \bibinfo {author} {\bibfnamefont
  {C.}~\bibnamefont {Cavazzoni}}, \bibinfo {author} {\bibfnamefont
  {D.}~\bibnamefont {Ceresoli}}, \bibinfo {author} {\bibfnamefont {G.~L.}\
  \bibnamefont {Chiarotti}}, \bibinfo {author} {\bibfnamefont {M.}~\bibnamefont
  {Cococcioni}}, \bibinfo {author} {\bibfnamefont {I.}~\bibnamefont {Dabo}},
  \bibinfo {author} {\bibfnamefont {A.~D.}\ \bibnamefont {Corso}}, \bibinfo
  {author} {\bibfnamefont {S.}~\bibnamefont {de~Gironcoli}}, \bibinfo {author}
  {\bibfnamefont {S.}~\bibnamefont {Fabris}}, \bibinfo {author} {\bibfnamefont
  {G.}~\bibnamefont {Fratesi}}, \bibinfo {author} {\bibfnamefont
  {R.}~\bibnamefont {Gebauer}}, \bibinfo {author} {\bibfnamefont
  {U.}~\bibnamefont {Gerstmann}}, \bibinfo {author} {\bibfnamefont
  {C.}~\bibnamefont {Gougoussis}}, \bibinfo {author} {\bibfnamefont
  {A.}~\bibnamefont {Kokalj}}, \bibinfo {author} {\bibfnamefont
  {M.}~\bibnamefont {Lazzeri}}, \bibinfo {author} {\bibfnamefont
  {L.}~\bibnamefont {Martin-Samos}}, \bibinfo {author} {\bibfnamefont
  {N.}~\bibnamefont {Marzari}}, \bibinfo {author} {\bibfnamefont
  {F.}~\bibnamefont {Mauri}}, \bibinfo {author} {\bibfnamefont
  {R.}~\bibnamefont {Mazzarello}}, \bibinfo {author} {\bibfnamefont
  {S.}~\bibnamefont {Paolini}}, \bibinfo {author} {\bibfnamefont
  {A.}~\bibnamefont {Pasquarello}}, \bibinfo {author} {\bibfnamefont
  {L.}~\bibnamefont {Paulatto}}, \bibinfo {author} {\bibfnamefont
  {C.}~\bibnamefont {Sbraccia}}, \bibinfo {author} {\bibfnamefont
  {S.}~\bibnamefont {Scandolo}}, \bibinfo {author} {\bibfnamefont
  {G.}~\bibnamefont {Sclauzero}}, \bibinfo {author} {\bibfnamefont {A.~P.}\
  \bibnamefont {Seitsonen}}, \bibinfo {author} {\bibfnamefont {A.}~\bibnamefont
  {Smogunov}}, \bibinfo {author} {\bibfnamefont {P.}~\bibnamefont {Umari}}, \
  and\ \bibinfo {author} {\bibfnamefont {R.~M.}\ \bibnamefont {Wentzcovitch}},\
  }\href {\doibase 10.1088/0953-8984/21/39/395502} {\bibfield  {journal}
  {\bibinfo  {journal} {Journal of Physics: Condensed Matter}\ }\textbf
  {\bibinfo {volume} {21}},\ \bibinfo {pages} {395502} (\bibinfo {year}
  {2009})}\BibitemShut {NoStop}%
\bibitem [{\citenamefont {Giannozzi}\ \emph {et~al.}(2017)\citenamefont
  {Giannozzi}, \citenamefont {Andreussi}, \citenamefont {Brumme}, \citenamefont
  {Bunau}, \citenamefont {Nardelli}, \citenamefont {Calandra}, \citenamefont
  {Car}, \citenamefont {Cavazzoni}, \citenamefont {Ceresoli}, \citenamefont
  {Cococcioni}, \citenamefont {Colonna}, \citenamefont {Carnimeo},
  \citenamefont {Corso}, \citenamefont {de~Gironcoli}, \citenamefont {Delugas},
  \citenamefont {Jr}, \citenamefont {Ferretti}, \citenamefont {Floris},
  \citenamefont {Fratesi}, \citenamefont {Fugallo}, \citenamefont {Gebauer},
  \citenamefont {Gerstmann}, \citenamefont {Giustino}, \citenamefont {Gorni},
  \citenamefont {Jia}, \citenamefont {Kawamura}, \citenamefont {Ko},
  \citenamefont {Kokalj}, \citenamefont {K\"u{\c c}\"ukbenli}, \citenamefont
  {Lazzeri}, \citenamefont {Marsili}, \citenamefont {Marzari}, \citenamefont
  {Mauri}, \citenamefont {Nguyen}, \citenamefont {Nguyen}, \citenamefont {de-la
  Roza}, \citenamefont {Paulatto}, \citenamefont {Ponc\'e}, \citenamefont
  {Rocca}, \citenamefont {Sabatini}, \citenamefont {Santra}, \citenamefont
  {Schlipf}, \citenamefont {Seitsonen}, \citenamefont {Smogunov}, \citenamefont
  {Timrov}, \citenamefont {Thonhauser}, \citenamefont {Umari}, \citenamefont
  {Vast}, \citenamefont {Wu},\ and\ \citenamefont {Baroni}}]{QE-2017}%
  \BibitemOpen
  \bibfield  {author} {\bibinfo {author} {\bibfnamefont {P.}~\bibnamefont
  {Giannozzi}}, \bibinfo {author} {\bibfnamefont {O.}~\bibnamefont
  {Andreussi}}, \bibinfo {author} {\bibfnamefont {T.}~\bibnamefont {Brumme}},
  \bibinfo {author} {\bibfnamefont {O.}~\bibnamefont {Bunau}}, \bibinfo
  {author} {\bibfnamefont {M.~B.}\ \bibnamefont {Nardelli}}, \bibinfo {author}
  {\bibfnamefont {M.}~\bibnamefont {Calandra}}, \bibinfo {author}
  {\bibfnamefont {R.}~\bibnamefont {Car}}, \bibinfo {author} {\bibfnamefont
  {C.}~\bibnamefont {Cavazzoni}}, \bibinfo {author} {\bibfnamefont
  {D.}~\bibnamefont {Ceresoli}}, \bibinfo {author} {\bibfnamefont
  {M.}~\bibnamefont {Cococcioni}}, \bibinfo {author} {\bibfnamefont
  {N.}~\bibnamefont {Colonna}}, \bibinfo {author} {\bibfnamefont
  {I.}~\bibnamefont {Carnimeo}}, \bibinfo {author} {\bibfnamefont {A.~D.}\
  \bibnamefont {Corso}}, \bibinfo {author} {\bibfnamefont {S.}~\bibnamefont
  {de~Gironcoli}}, \bibinfo {author} {\bibfnamefont {P.}~\bibnamefont
  {Delugas}}, \bibinfo {author} {\bibfnamefont {R.~A.~D.}\ \bibnamefont {Jr}},
  \bibinfo {author} {\bibfnamefont {A.}~\bibnamefont {Ferretti}}, \bibinfo
  {author} {\bibfnamefont {A.}~\bibnamefont {Floris}}, \bibinfo {author}
  {\bibfnamefont {G.}~\bibnamefont {Fratesi}}, \bibinfo {author} {\bibfnamefont
  {G.}~\bibnamefont {Fugallo}}, \bibinfo {author} {\bibfnamefont
  {R.}~\bibnamefont {Gebauer}}, \bibinfo {author} {\bibfnamefont
  {U.}~\bibnamefont {Gerstmann}}, \bibinfo {author} {\bibfnamefont
  {F.}~\bibnamefont {Giustino}}, \bibinfo {author} {\bibfnamefont
  {T.}~\bibnamefont {Gorni}}, \bibinfo {author} {\bibfnamefont
  {J.}~\bibnamefont {Jia}}, \bibinfo {author} {\bibfnamefont {M.}~\bibnamefont
  {Kawamura}}, \bibinfo {author} {\bibfnamefont {H.-Y.}\ \bibnamefont {Ko}},
  \bibinfo {author} {\bibfnamefont {A.}~\bibnamefont {Kokalj}}, \bibinfo
  {author} {\bibfnamefont {E.}~\bibnamefont {K\"u{\c c}\"ukbenli}}, \bibinfo
  {author} {\bibfnamefont {M.}~\bibnamefont {Lazzeri}}, \bibinfo {author}
  {\bibfnamefont {M.}~\bibnamefont {Marsili}}, \bibinfo {author} {\bibfnamefont
  {N.}~\bibnamefont {Marzari}}, \bibinfo {author} {\bibfnamefont
  {F.}~\bibnamefont {Mauri}}, \bibinfo {author} {\bibfnamefont {N.~L.}\
  \bibnamefont {Nguyen}}, \bibinfo {author} {\bibfnamefont {H.-V.}\
  \bibnamefont {Nguyen}}, \bibinfo {author} {\bibfnamefont {A.~O.}\
  \bibnamefont {de-la Roza}}, \bibinfo {author} {\bibfnamefont
  {L.}~\bibnamefont {Paulatto}}, \bibinfo {author} {\bibfnamefont
  {S.}~\bibnamefont {Ponc\'e}}, \bibinfo {author} {\bibfnamefont
  {D.}~\bibnamefont {Rocca}}, \bibinfo {author} {\bibfnamefont
  {R.}~\bibnamefont {Sabatini}}, \bibinfo {author} {\bibfnamefont
  {B.}~\bibnamefont {Santra}}, \bibinfo {author} {\bibfnamefont
  {M.}~\bibnamefont {Schlipf}}, \bibinfo {author} {\bibfnamefont {A.~P.}\
  \bibnamefont {Seitsonen}}, \bibinfo {author} {\bibfnamefont {A.}~\bibnamefont
  {Smogunov}}, \bibinfo {author} {\bibfnamefont {I.}~\bibnamefont {Timrov}},
  \bibinfo {author} {\bibfnamefont {T.}~\bibnamefont {Thonhauser}}, \bibinfo
  {author} {\bibfnamefont {P.}~\bibnamefont {Umari}}, \bibinfo {author}
  {\bibfnamefont {N.}~\bibnamefont {Vast}}, \bibinfo {author} {\bibfnamefont
  {X.}~\bibnamefont {Wu}}, \ and\ \bibinfo {author} {\bibfnamefont
  {S.}~\bibnamefont {Baroni}},\ }\href
  {http://stacks.iop.org/0953-8984/29/i=46/a=465901} {\bibfield  {journal}
  {\bibinfo  {journal} {Journal of Physics: Condensed Matter}\ }\textbf
  {\bibinfo {volume} {29}},\ \bibinfo {pages} {465901} (\bibinfo {year}
  {2017})}\BibitemShut {NoStop}%
\bibitem [{\citenamefont {Kresse}\ and\ \citenamefont
  {Hafner}(1994{\natexlab{a}})}]{Kresse1994}%
  \BibitemOpen
  \bibfield  {author} {\bibinfo {author} {\bibfnamefont {G.}~\bibnamefont
  {Kresse}}\ and\ \bibinfo {author} {\bibfnamefont {J.}~\bibnamefont
  {Hafner}},\ }\href {\doibase 10.1088/0953-8984/6/40/015} {\bibfield
  {journal} {\bibinfo  {journal} {Journal of Physics: Condensed Matter}\
  }\textbf {\bibinfo {volume} {6}},\ \bibinfo {pages} {8245} (\bibinfo {year}
  {1994}{\natexlab{a}})}\BibitemShut {NoStop}%
\bibitem [{\citenamefont {Gygi}\ and\ \citenamefont
  {Baldereschi}(1986)}]{Gygi-1986}%
  \BibitemOpen
  \bibfield  {author} {\bibinfo {author} {\bibfnamefont {F.}~\bibnamefont
  {Gygi}}\ and\ \bibinfo {author} {\bibfnamefont {A.}~\bibnamefont
  {Baldereschi}},\ }\href {\doibase 10.1103/PhysRevB.34.4405} {\bibfield
  {journal} {\bibinfo  {journal} {Phys. Rev. B}\ }\textbf {\bibinfo {volume}
  {34}},\ \bibinfo {pages} {4405} (\bibinfo {year} {1986})}\BibitemShut
  {NoStop}%
\bibitem [{\citenamefont {\"Ostlund}\ and\ \citenamefont
  {Rommer}(1995)}]{Ostlund-1995}%
  \BibitemOpen
  \bibfield  {author} {\bibinfo {author} {\bibfnamefont {S.}~\bibnamefont
  {\"Ostlund}}\ and\ \bibinfo {author} {\bibfnamefont {S.}~\bibnamefont
  {Rommer}},\ }\href {\doibase 10.1103/PhysRevLett.75.3537} {\bibfield
  {journal} {\bibinfo  {journal} {Phys. Rev. Lett.}\ }\textbf {\bibinfo
  {volume} {75}},\ \bibinfo {pages} {3537} (\bibinfo {year}
  {1995})}\BibitemShut {NoStop}%
\bibitem [{\citenamefont {Barcza}\ \emph {et~al.}(2013)\citenamefont {Barcza},
  \citenamefont {Barford}, \citenamefont {Gebhard},\ and\ \citenamefont
  {Legeza}}]{Barcza-2013}%
  \BibitemOpen
  \bibfield  {author} {\bibinfo {author} {\bibfnamefont {G.}~\bibnamefont
  {Barcza}}, \bibinfo {author} {\bibfnamefont {W.}~\bibnamefont {Barford}},
  \bibinfo {author} {\bibfnamefont {F.}~\bibnamefont {Gebhard}}, \ and\
  \bibinfo {author} {\bibfnamefont {O.}~\bibnamefont {Legeza}},\ }\href
  {\doibase 10.1103/PhysRevB.87.245116} {\bibfield  {journal} {\bibinfo
  {journal} {Phys. Rev. B}\ }\textbf {\bibinfo {volume} {87}},\ \bibinfo
  {pages} {245116} (\bibinfo {year} {2013})}\BibitemShut {NoStop}%
\bibitem [{\citenamefont {Hu}\ and\ \citenamefont {Chan}(2015)}]{Hu-2015}%
  \BibitemOpen
  \bibfield  {author} {\bibinfo {author} {\bibfnamefont {W.}~\bibnamefont
  {Hu}}\ and\ \bibinfo {author} {\bibfnamefont {G.~K.-L.}\ \bibnamefont
  {Chan}},\ }\href {\doibase 10.1021/acs.jctc.5b00174} {\bibfield  {journal}
  {\bibinfo  {journal} {Journal of Chemical Theory and Computation}\ }\textbf
  {\bibinfo {volume} {11}},\ \bibinfo {pages} {3000} (\bibinfo {year}
  {2015})}\BibitemShut {NoStop}%
\bibitem [{\citenamefont {Bockstedte}\ \emph {et~al.}(2018)\citenamefont
  {Bockstedte}, \citenamefont {Sch{\"u}tz}, \citenamefont {Garratt},
  \citenamefont {Iv{\'a}dy},\ and\ \citenamefont {Gali}}]{BockstedteNPJ2018}%
  \BibitemOpen
  \bibfield  {author} {\bibinfo {author} {\bibfnamefont {M.}~\bibnamefont
  {Bockstedte}}, \bibinfo {author} {\bibfnamefont {F.}~\bibnamefont
  {Sch{\"u}tz}}, \bibinfo {author} {\bibfnamefont {T.}~\bibnamefont {Garratt}},
  \bibinfo {author} {\bibfnamefont {V.}~\bibnamefont {Iv{\'a}dy}}, \ and\
  \bibinfo {author} {\bibfnamefont {A.}~\bibnamefont {Gali}},\ }\href {\doibase
  10.1038/s41535-018-0103-6} {\bibfield  {journal} {\bibinfo  {journal} {npj
  Quantum Materials}\ }\textbf {\bibinfo {volume} {3}},\ \bibinfo {pages} {31}
  (\bibinfo {year} {2018})}\BibitemShut {NoStop}%
\bibitem [{\citenamefont {Stein}\ and\ \citenamefont
  {Reiher}(2016)}]{Stein2016}%
  \BibitemOpen
  \bibfield  {author} {\bibinfo {author} {\bibfnamefont {C.~J.}\ \bibnamefont
  {Stein}}\ and\ \bibinfo {author} {\bibfnamefont {M.}~\bibnamefont {Reiher}},\
  }\href {\doibase 10.1021/acs.jctc.6b00156} {\bibfield  {journal} {\bibinfo
  {journal} {Journal of Chemical Theory and Computation}\ }\textbf {\bibinfo
  {volume} {12}},\ \bibinfo {pages} {1760} (\bibinfo {year}
  {2016})}\BibitemShut {NoStop}%
\bibitem [{\citenamefont {Zhu}\ \emph {et~al.}(2020)\citenamefont {Zhu},
  \citenamefont {Cui},\ and\ \citenamefont {Chan}}]{Zhu2020}%
  \BibitemOpen
  \bibfield  {author} {\bibinfo {author} {\bibfnamefont {T.}~\bibnamefont
  {Zhu}}, \bibinfo {author} {\bibfnamefont {Z.-H.}\ \bibnamefont {Cui}}, \ and\
  \bibinfo {author} {\bibfnamefont {G.~K.-L.}\ \bibnamefont {Chan}},\ }\href
  {https://doi.org/10.1021/acs.jctc.9b00934} {\bibfield  {journal} {\bibinfo
  {journal} {Journal of Chemical Theory and Computation}\ }\textbf {\bibinfo
  {volume} {16}},\ \bibinfo {pages} {141} (\bibinfo {year} {2020})}\BibitemShut
  {NoStop}%
\bibitem [{\citenamefont {Sajid}\ \emph {et~al.}(2020)\citenamefont {Sajid},
  \citenamefont {Ford},\ and\ \citenamefont {Reimers}}]{Sajid-2019a}%
  \BibitemOpen
  \bibfield  {author} {\bibinfo {author} {\bibfnamefont {A.}~\bibnamefont
  {Sajid}}, \bibinfo {author} {\bibfnamefont {M.~J.}\ \bibnamefont {Ford}}, \
  and\ \bibinfo {author} {\bibfnamefont {J.~R.}\ \bibnamefont {Reimers}},\
  }\href {\doibase 10.1088/1361-6633/ab6310} {\bibfield  {journal} {\bibinfo
  {journal} {Reports on Progress in Physics}\ }\textbf {\bibinfo {volume}
  {83}},\ \bibinfo {pages} {044501} (\bibinfo {year} {2020})}\BibitemShut
  {NoStop}%
\bibitem [{\citenamefont {Perdew}\ \emph {et~al.}(1996)\citenamefont {Perdew},
  \citenamefont {Burke},\ and\ \citenamefont {Ernzerhof}}]{PBE}%
  \BibitemOpen
  \bibfield  {author} {\bibinfo {author} {\bibfnamefont {J.~P.}\ \bibnamefont
  {Perdew}}, \bibinfo {author} {\bibfnamefont {K.}~\bibnamefont {Burke}}, \
  and\ \bibinfo {author} {\bibfnamefont {M.}~\bibnamefont {Ernzerhof}},\ }\href
  {\doibase 10.1103/PhysRevLett.77.3865} {\bibfield  {journal} {\bibinfo
  {journal} {Phys. Rev. Lett.}\ }\textbf {\bibinfo {volume} {77}},\ \bibinfo
  {pages} {3865} (\bibinfo {year} {1996})}\BibitemShut {NoStop}%
\bibitem [{\citenamefont {Kresse}\ and\ \citenamefont
  {Hafner}(1994{\natexlab{b}})}]{VASP}%
  \BibitemOpen
  \bibfield  {author} {\bibinfo {author} {\bibfnamefont {G.}~\bibnamefont
  {Kresse}}\ and\ \bibinfo {author} {\bibfnamefont {J.}~\bibnamefont
  {Hafner}},\ }\href {\doibase 10.1103/PhysRevB.49.14251} {\bibfield  {journal}
  {\bibinfo  {journal} {Phys. Rev. B}\ }\textbf {\bibinfo {volume} {49}},\
  \bibinfo {pages} {14251} (\bibinfo {year} {1994}{\natexlab{b}})}\BibitemShut
  {NoStop}%
\bibitem [{\citenamefont {Neese}(2012)}]{Neese-2012}%
  \BibitemOpen
  \bibfield  {author} {\bibinfo {author} {\bibfnamefont {F.}~\bibnamefont
  {Neese}},\ }\href {\doibase 10.1002/wcms.81} {\bibfield  {journal} {\bibinfo
  {journal} {WIREs Computational Molecular Science}\ }\textbf {\bibinfo
  {volume} {2}},\ \bibinfo {pages} {73} (\bibinfo {year} {2012})}\BibitemShut
  {NoStop}%
\bibitem [{\citenamefont {Kallay}\ \emph
  {et~al.}(2020{\natexlab{a}})\citenamefont {Kallay}, \citenamefont {Nagy},
  \citenamefont {Mester}, \citenamefont {Rolik}, \citenamefont {Samu},
  \citenamefont {Csontos}, \citenamefont {Csoka}, \citenamefont {Szabo},
  \citenamefont {Gyevi-Nagy}, \citenamefont {Hegely}, \citenamefont
  {Ladjanszki}, \citenamefont {Szegedy}, \citenamefont {Ladoczki},
  \citenamefont {Petrov}, \citenamefont {Farkas}, \citenamefont {Mezei},\ and\
  \citenamefont {Ganyecz}}]{mrcc2020}%
  \BibitemOpen
  \bibfield  {author} {\bibinfo {author} {\bibfnamefont {M.}~\bibnamefont
  {Kallay}}, \bibinfo {author} {\bibfnamefont {P.~R.}\ \bibnamefont {Nagy}},
  \bibinfo {author} {\bibfnamefont {D.}~\bibnamefont {Mester}}, \bibinfo
  {author} {\bibfnamefont {Z.}~\bibnamefont {Rolik}}, \bibinfo {author}
  {\bibfnamefont {G.}~\bibnamefont {Samu}}, \bibinfo {author} {\bibfnamefont
  {J.}~\bibnamefont {Csontos}}, \bibinfo {author} {\bibfnamefont
  {J.}~\bibnamefont {Csoka}}, \bibinfo {author} {\bibfnamefont {B.~P.}\
  \bibnamefont {Szabo}}, \bibinfo {author} {\bibfnamefont {L.}~\bibnamefont
  {Gyevi-Nagy}}, \bibinfo {author} {\bibfnamefont {B.}~\bibnamefont {Hegely}},
  \bibinfo {author} {\bibfnamefont {I.}~\bibnamefont {Ladjanszki}}, \bibinfo
  {author} {\bibfnamefont {L.}~\bibnamefont {Szegedy}}, \bibinfo {author}
  {\bibfnamefont {B.}~\bibnamefont {Ladoczki}}, \bibinfo {author}
  {\bibfnamefont {K.}~\bibnamefont {Petrov}}, \bibinfo {author} {\bibfnamefont
  {M.}~\bibnamefont {Farkas}}, \bibinfo {author} {\bibfnamefont {P.~D.}\
  \bibnamefont {Mezei}}, \ and\ \bibinfo {author} {\bibfnamefont
  {A.}~\bibnamefont {Ganyecz}},\ }\href {www.mrcc.hu} {\enquote {\bibinfo
  {title} {Mrcc, a quantum chemical program suite},}\ } (\bibinfo {year}
  {2020}{\natexlab{a}})\BibitemShut {NoStop}%
\bibitem [{\citenamefont {Kallay}\ \emph
  {et~al.}(2020{\natexlab{b}})\citenamefont {Kallay}, \citenamefont {Nagy},
  \citenamefont {Mester}, \citenamefont {Rolik}, \citenamefont {Samu},
  \citenamefont {Csontos}, \citenamefont {Csoka}, \citenamefont {Szabo},
  \citenamefont {Gyevi-Nagy}, \citenamefont {Hegely}, \citenamefont
  {Ladjanszki}, \citenamefont {Szegedy}, \citenamefont {Ladoczki},
  \citenamefont {Petrov}, \citenamefont {Farkas}, \citenamefont {Mezei},\ and\
  \citenamefont {Ganyecz}}]{mrcc2020b}%
  \BibitemOpen
  \bibfield  {author} {\bibinfo {author} {\bibfnamefont {M.}~\bibnamefont
  {Kallay}}, \bibinfo {author} {\bibfnamefont {P.~R.}\ \bibnamefont {Nagy}},
  \bibinfo {author} {\bibfnamefont {D.}~\bibnamefont {Mester}}, \bibinfo
  {author} {\bibfnamefont {Z.}~\bibnamefont {Rolik}}, \bibinfo {author}
  {\bibfnamefont {G.}~\bibnamefont {Samu}}, \bibinfo {author} {\bibfnamefont
  {J.}~\bibnamefont {Csontos}}, \bibinfo {author} {\bibfnamefont
  {J.}~\bibnamefont {Csoka}}, \bibinfo {author} {\bibfnamefont
  {B.}~\bibnamefont {Szabo}}, \bibinfo {author} {\bibfnamefont
  {L.}~\bibnamefont {Gyevi-Nagy}}, \bibinfo {author} {\bibfnamefont
  {B.}~\bibnamefont {Hegely}}, \bibinfo {author} {\bibfnamefont
  {I.}~\bibnamefont {Ladjanszki}}, \bibinfo {author} {\bibfnamefont
  {L.}~\bibnamefont {Szegedy}}, \bibinfo {author} {\bibfnamefont
  {B.}~\bibnamefont {Ladoczki}}, \bibinfo {author} {\bibfnamefont
  {K.}~\bibnamefont {Petrov}}, \bibinfo {author} {\bibfnamefont
  {M.}~\bibnamefont {Farkas}}, \bibinfo {author} {\bibfnamefont
  {P.}~\bibnamefont {Mezei}}, \ and\ \bibinfo {author} {\bibfnamefont
  {A.}~\bibnamefont {Ganyecz}},\ }\href {\doibase 10.1063/1.5142048} {\bibfield
   {journal} {\bibinfo  {journal} {The Journal of Chemical Physics}\ }\textbf
  {\bibinfo {volume} {152}},\ \bibinfo {pages} {074107} (\bibinfo {year}
  {2020}{\natexlab{b}})}\BibitemShut {NoStop}%
\bibitem [{\citenamefont {Legeza}\ \emph {et~al.}(2020)\citenamefont {Legeza},
  \citenamefont {Veis},\ and\ \citenamefont {Mosoni}}]{budapest_qcdmrg}%
  \BibitemOpen
  \bibfield  {author} {\bibinfo {author} {\bibfnamefont {{\"O}.}~\bibnamefont
  {Legeza}}, \bibinfo {author} {\bibfnamefont {L.}~\bibnamefont {Veis}}, \ and\
  \bibinfo {author} {\bibfnamefont {T.}~\bibnamefont {Mosoni}},\ }\href@noop {}
  {\enquote {\bibinfo {title} {{QC-DMRG-Budapest, a program for quantum
  chemical DMRG calculations}},}\ } (\bibinfo {year} {2020})\BibitemShut
  {NoStop}%
\end{thebibliography}

%

\end{document}